\documentclass{article}

\usepackage{PRIMEarxiv}

\usepackage[utf8]{inputenc} % allow utf-8 input
\usepackage[T1]{fontenc}    % use 8-bit T1 fonts
\usepackage[hidelinks]{hyperref}
\usepackage{url}            % simple URL typesetting
\usepackage{booktabs}       % professional-quality tables
\usepackage{amsfonts}       % blackboard math symbols
\usepackage{nicefrac}       % compact symbols for 1/2, etc.
\usepackage{microtype}      % microtypography
\usepackage{lipsum}
\usepackage{fancyhdr}       % header
\usepackage{graphicx}       % graphics
\graphicspath{{media/}}     % organize your images and other figures under media/ folder

\usepackage[toc,page]{appendix}

\usepackage{graphicx}
\usepackage{multirow}
\usepackage{multicol}
\usepackage{amsmath}
\usepackage{acronym}
\usepackage{mdframed}
\usepackage{pdfpages}
\usepackage{framed}
\usepackage{tikz}
\usepackage{xspace}
\usepackage{enumitem}

\usetikzlibrary{shapes.geometric, arrows}
\usetikzlibrary{trees,positioning,shapes,shadows,arrows}
\usepackage{lipsum}
\usepackage{pgfplots} 
\usepackage{forest}
\usepackage{caption}
\usepackage{amssymb}
\usepackage{xcolor}
\usepackage{tabularray}
\usepackage{setspace}
\pgfplotsset{compat=1.18}
\acrodef{IFT}{Incident Fault Tree}
\acrodef{CE}{Cyber Essentials}
\acrodef{TTP}{Tactic Techniques and Procedures}
\acrodef{AC}{Additional Controls}
\acrodef{VE}{Vulnerability Exploitation}
\newenvironment{finding}{\begin{framed}\textbf{Take Away.}\noindent~}{\end{framed}}

\title{Assessing Effectiveness of Cyber Essentials Technical Controls
}

\author{
  Priyanka Badva, Partha Das Chowdhury, Kopo M. Ramokapane, Barnaby Craggs, Awais Rashid \\
  \\
  University of Bristol \\
  UK \\
}

\begin{document}
\maketitle

\linespread{3}

\begin{abstract}
Cyber Essentials (CE) comprise a set of controls designed to protect organisations, irrespective of their size, against cyber attacks. The controls are \emph{firewalls, secure configuration, user access control, malware protection} \& \emph{security update management}. In this work, we explore the extent to which CE remains robust against an ever-evolving threat landscape. To that end, we reconstruct 45 breaches mapped to MiTRE ATT\&CK using an \ac{IFT} approach. Our method reveals the intersections where the placement of controls could have protected organisations. Then we identify appropriate Cyber Essential controls and/or Additional Controls for these vulnerable intersections. Our results show that CE controls can effectively protect against most attacks during the initial attack phase. However, they may need to be complemented with Additional Controls if the attack proceeds further into organisational systems \& networks. The Additional Controls (AC) we identify include back-ups, security awareness, logging and monitoring. Our analysis brings to the fore a foundational issue as to whether controls should exclude recovery and focus only on pre-emption. The latter makes the strong assumption that a prior identification of all controls in a dynamic threat landscape is indeed possible. Furthermore, any potential broadening of technical controls entails re-scoping the skills that are required for a \ac{CE} assessor. To that end, we suggest \emph{human factors} and \emph{security operations and incident management} as two potential knowledge areas from Cyber Security Body of Knowledge (CyBOK) if there is any broadening of \ac{CE} based on these findings.
\end{abstract}

% keywords can be removed

\section{Introduction}
The Cyber Essentials (\ac{CE}) technical controls~\cite{essentials2022requirements} are the UK government certification scheme to help organisations secure themselves against attacks on their information systems and processes. The framework comprises a set of technical controls developed in collaboration with the Information Assurance for Small and Medium Enterprises (IASME) consortium, the Information Security Forum (ISF) and the British Standards Institution (BSI). \ac{CE} certification is a pre-requisite to bid for UK government contracts, particularly in areas that involve handling sensitive information. A study~\cite{lancasterreport} conducted in 2015 to assess the effectiveness of Cyber Essential technical controls 
% conclude that they are effective in mitigating attacks to information systems and networks. The study 
investigated 200 vulnerabilities from the CVE list of 2013 and 2014 published by MiTRE. The study  reported that 131 vulnerabilities were mitigated using \ac{CE} technical controls. \ac{CE} controls were also reported to be effective against three high-profile vulnerabilities of the day, namely Heartbleed, ShellShock and Superfish. 

The landscape of threats and mal-actors has changed over 8 years since that first systematic evaluation of \ac{CE}. The proliferation of smartphones and clouds has increasingly replaced conventional data stores holding sensitive information. Compromised email accounts cost Yahoo to the tune of \$350 million~\cite{anderson2013,anderson2019measuring}. % During 2017 malware developed at the highest levels of US government was used to develop WannaCry, Petya/NotPetya ransomware~\cite{fortinet}. 
WannaCry led to disruptions in the National Health Service (NHS) information systems~\cite{wannacry} and NotPetya spread through an update to the MEDoc tax accounting software~\cite{notpetya}. Cryptocurrencies, along with an illicit market of exploits, facilitate ransomware as a service~\cite{karapapas2020ransomware,pastrana2019first} to low-skilled criminal gangs. 

Motivated by the previous study~\cite{lancasterreport} and the changing threat landscape, we explore the extent to which \ac{CE} controls remain robust as a basic cyber hygiene mechanism. We do so through a systematic analysis of 45 incident reports collected through a research agreement with a managed security services provider. These attacks are mapped to the MiTRE ATT\&CK knowledge base and include a set of ransomware incidents. Our goal was to understand the extent to which \ac{CE} controls could have prevented those breaches. The Research Questions (RQs) we address in this work as follows:\\

\noindent \textbf{RQ1.} Which attack pathways could have been  inhibited by \ac{CE} technical controls and, to what extent, i.e., whether these controls would completely  inhibit those pathways or whether layering with other controls (multiple \ac{CE} controls or a wider set) would have been required?\\
\textbf{RQ2.} Which controls beyond \ac{CE} appear frequently as  inhibitors of attacks and how effective are they in  inhibiting particular attack pathways (on their own or in conjunction with other – including \ac{CE} – controls)?\\
\textbf{RQ3.} Are there specific patterns of  inhibition (whether \ac{CE} controls or others) with regards to the subset of ransomware incidents in the corpus?\\

We used the Incident Fault Tree (\ac{IFT}) approach ~\cite{rashidift,johnson2003handbook} to reconstruct the attacks in terms of events and mitigation mechanisms that could have prevented them. The method is an adaptation of the method used in our prior work on analysing data exfiltration attacks~\cite{rashidift}. It enables systematic modelling of attacks and, critically, the points at which specific security controls provide suitable mitigations against particular attack pathways. In our model of the attacks, the mitigation mechanisms are expressed as \emph{inhibit gates}. The presence or absence of \emph{inhibit gates} in comparison to the previous reports are analysed to understand the resilience of \ac{CE} across space (i.e., the classes of attacks and types of attack pathways as modelled using MiTRE ATT\&CK) and time (change in malware and attack behaviours since the initial research in those reports were conducted). For example, prior analysis shows an overwhelming majority of the breaches (131/200) were completely mitigated through Cyber Essential controls.

Our analysis shows that \ac{CE} controls by themselves effectively mitigate most of the attack scenarios at the initial stages of the attack lifecycle. For instance, we find from our investigations of the ransomware incidents that \ac{CE} controls can stop them at their initial stages. However, in some cases, as the attacks progress deeper into systems and networks, \ac{CE} controls must be complemented with \ac{AC}. The pertinent question then is, whether we should consider incorporating \ac{AC} controls as part of \ac{CE}. This entails broadening the definition of controls to include recovery mechanisms such as back-ups. Finally, we synthesised our findings with the mapping of the National Cyber Security Centre (NCSC) Cyber Advisor \ac{CE} criterion to the Cyber Security Body of Knowledge (CyBOK)~\cite{cybok}. This mapping shows that the assessors are expected to have substantial knowledge of \emph{cloud security} and \emph{applied cryptography}. This may be an indicator of the changing security needs over space and time. Expanding on the argument of broadening the scope of mitigation mechanisms, the \ac{AC} we find as part of our investigations point towards the need for assessors to be conversant in additional knowledge areas like \emph{Human Factors} and \emph{Security Operations and Incident Management}. Synthesising this mapping with the analysis from our \ac{IFT}s and the cross-contrast with prior work enables us to build a fuller picture of the effectiveness of the current controls and also whether \ac{AC} merit recommendation for inclusion. % (as well as evidence to back any such recommendation).

We draw upon related work in Section \ref{related work} while discussing our contribution. In Section \ref{background} we briefly discuss \ac{CE} controls, the constituent elements and notations of \ac{IFT} and MiTRE ATT\&CK framework. This is followed by an overview of the dataset of the attacks in Section~\ref{dataset}. Section~\ref{method} details the methods we use to analyse and arrive at the mitigation mechanisms. In Section~\ref{analysis}, we outline the \emph{edge}, \emph{level}, \emph{phase} and \emph{level+phase} analysis we use on the dataset which is followed by our findings in Section~\ref{findings}. We discuss some of the key findings like \ac{AC}, situation of the controls and compare with prior validation exercises in Section~\ref{discussion} which is followed by conclusion in Section~\ref{conclusion}.

\section{Related Work}\label{related work}
Such et al. discuss the effectiveness of basic cyber hygiene practices in preventing cyber attacks~\cite{such2019basic}. The term ``basic cyber hygiene" refers to a set of fundamental security practices recommended by many experts as a starting point for organisations to improve their cyber security posture. The paper reviews several studies and surveys that have examined the impact of basic cyber hygiene practices on cyber security outcomes. The studies suggest that organisations that implement basic cyber hygiene practices, such as keeping software up to date, using strong passwords, and regularly backing up data, are less likely to experience cyber attacks and data breaches. The authors also note that, while basic cyber hygiene practices can be effective, they are not a silver bullet solution for cyber security. Sophisticated attackers can still find ways to bypass these controls, and organisations should not rely solely on basic cyber hygiene practices to protect themselves.

The article published by Lancaster University ``Cyber Security Controls Effectiveness: A Qualitative Assessment of Cyber Essentials" discusses the effectiveness of the \ac{CE} controls~\cite{lancasterreport}. The authors conducted a qualitative assessment of \ac{CE} by interviewing cyber security experts and practitioners from a range of industries. They asked the participants to rate the effectiveness of the \ac{CE} controls in preventing various types of cyber attacks, such as phishing, malware, and insider threats. The study found that while the \ac{CE} controls were generally effective at preventing common cyber threats, there were some limitations to their effectiveness. The authors also identified several factors that could impact the effectiveness of the \ac{CE} controls, such as the size and complexity of an organisation's IT environment, the skill and expertise of its IT staff, and the level of employee awareness and training. Overall, the study suggests that organisations should view \ac{CE} as a starting point for improving their cyber security and should continue to assess and improve their security posture over time.

The findings of~\cite{such2019basic} and~\cite{lancasterreport} both point towards the need for \ac{AC} in light of the changing threat landscape. For example~\cite{lancasterreport}, highlights the need for awareness and monitoring among others. We contribute to this existing body of work as: 
\begin{itemize}
    \item In relation to prior studies, we find that \ac{CE}  controls are effective against a wide range of cyber attacks as has been reported in~\cite{such2019basic} \&~\cite{lancasterreport}. We find that patch management is not as prominent as a control in our study compared to prior evaluations~\cite{such2019basic}. Similar to prior evaluations we identify the need for \ac{AC}. Some of the \ac{AC} controls we propose are the same as~\cite{lancasterreport} while we find the need for \emph{back up} and \emph{encryption} to mitigate some of the attack pathways in our corpus of incidents. In addition to the secure design defaults as noted by~\cite{lancasterreport} we identify the need for secure deployment defaults to mitigate many attack pathways at their initial stages. 
    \item  Compared to prior studies~\cite{such2019basic} \&~\cite{lancasterreport}, we additionally focussed on ransomware attacks from our corpus of incidents. In these cases, \ac{CE} controls by themselves can be applied to mitigate most of the attack pathways, however they need support of \ac{AC} in many instances.  
    \item Beyond recommending \ac{AC}, we raise an issue which is fairly foundational to the purpose of \ac{CE}. Should controls only be preventive in nature or also pre-emptive? The former requires a complete understanding of all the possible threats well ahead of time. For business continuity it is important that systems respond to attacks as they happen or are imminent. To that end we suggest \ac{AC} such as \emph{back up} that can pre-empt imminent or in progress attacks, for example, ransomware attacks. 
\end{itemize}

\section{Background}\label{background}

\subsection{Cyber Essentials Technical Controls}

There are five technical controls organised under Cyber Essentials requirement for IT infrastructures\footnote{We use Version 3.0 as that was the current version at the time of the analysis.}~\cite{essentials2022requirements}:
\begin{itemize}
    \item \textbf{Firewall:} Firewalls mediate between traffic that leaves and enters a network through policies. The policies protect against unauthorised users getting inside a network and such policies can be specified in terms of source (who can access the network), destination (where users within the network can visit) and the type of services that are allowed in a network. 
    \item \textbf{Secure Configuration:} Systems come with a default configuration, for example, default admin user name and passwords which, if not changed, can allow attackers to exploit the system to harm the legitimate users. 
    \item \textbf{User Access Control:} Organisations should link the stable identity of their legitimate users to resources to prevent unauthorised access to the resources. The legitimate users would in turn prove their right to access such resources through something they know or they possess. This control mandates organisations should carefully onboard users, securely maintain their user list and remove users once they leave the organisation. This has implications on the security of other controls. Moreover, an organisation should also implement Two-Factor Authentication (2FA) as an additional layer of security, which helps in  reducing the risk of unauthorised user access and enhances the overall security of access control systems.
    \item \textbf{Malware Protection:} Systems inside an organisation's network should not be readily susceptible to malicious pieces of code. The malicious code can come from trusted sources as well. So there need to be protections to mitigate malware entering the organisation's systems and propagating. 
    \item \textbf{Security Update Management:} Bugs do emerge in software systems much later than they are released and vendors provide updates to the existing insecure version. While the internet makes it easier to remotely patch insecure software, organisations need the discipline to apply such updates regularly. 
\end{itemize}

\subsection{Incident Fault Tree}
An Incident Fault Tree \ac{IFT}~\cite{johnson2003handbook} is a modelling approach to explain incidents in terms of various causes and events. The method proposed in~\cite{rashidift} involves modelling each security incident as an \ac{IFT}. An \ac{IFT} is a modelling tool for the retrospective documentation and analysis of an incident. It is used to explain the incident in terms of the various causes that contributed to it. An \ac{IFT} is based on, and thus uses, the graphical symbols of the Fault Tree Analysis (FTA) method. It comprises two broad types of notations, namely \emph{events} and \emph{gates}. \emph{Events} are specific discrete events or other conditions that precipitate an incident. \emph{Gates} are used to represent multiple scenarios. They can be used to represent the various ways in which \emph{events} combine in order to cause another event. The mitigating factors that can prevent an exploitable event are also represented as \emph{ inhibit gates}. The \ac{IFT} symbols are shown in Figure~\ref{fig:ift_symbols} and are as follows: %As shown in Fig. 1, an \ac{IFT} has the following elements (an example \ac{IFT} of the Nitro attacks is shown in (Fig. 2)): 
\begin{figure}[ht]
    \centering
    \includegraphics[scale=0.30]{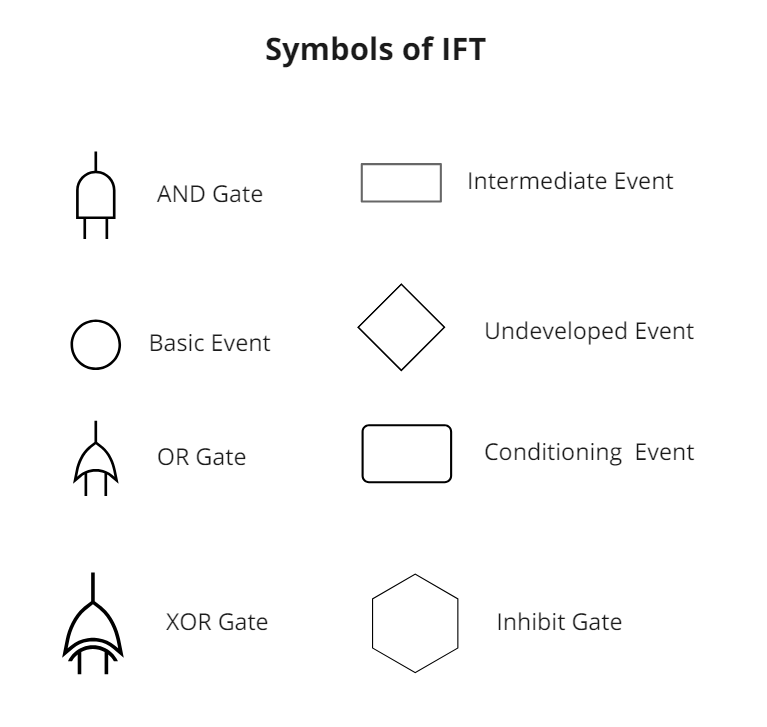}
    \caption{IFT Symbols}
    \label{fig:ift_symbols}
\end{figure}

\begin{itemize}
    \item Basic Event: cannot be explained in terms of other causal events, or has no known causes;
    \item Intermediate Event: can be explained in terms of other causal events, or there are known causes;
 \item Undeveloped Event: can be explained in terms of other causal events but we have chosen not to do so because such causal elaboration serves no analytical purpose;
 \item AND Gate and OR Gate: respectively represent the conjunction or disjunction of causal events;
 \item  Inhibit Gate: is placed between an event and its causes to stop an event from occurring unless the condition in the Conditioning Event is satisfied. The Inhibit Gate is a critical part of the modelling and analysis as it makes it possible to model the mitigating or remedial actions that may act as barriers to the progress towards a security breach. In this project, we use it to model where \ac{CE} Controls would have inhibited an attack and where other controls beyond \ac{CE} were required.
\end{itemize}

\subsection{MiTRE ATT\&CK }

The behaviour and attack paths of an adversary for known attacks are documented in the Adversarial Tactics Techniques \& Common Knowledge (ATT\&CK) framework~\cite{strom2018MiTRE}. The project started with documenting the \ac{TTP} of adversaries attacking Microsoft Windows systems which gradually expanded to a general record of adversarial behaviour on other platforms such as Linux, macOS, mobile devices and industrial control systems. This body of knowledge details pre \& post compromise behaviour and is a useful tool for organisations to understand the potential vulnerable elements in their information and communication infrastructure.  

At its core, ATT\&CK is a behavioural model that consists of multiple core components:

\begin{itemize}
    \item Firstly, tactics refer to the short-term tactical goals adversaries aim to achieve during an attack.
    
    \item Secondly, techniques describe the methods that adversaries use to achieve their tactical goals. These methods may include exploiting vulnerabilities, social engineering, or malicious software.
    
    \item Thirdly, sub-techniques provide more specific details on the means by which adversaries achieve their tactical goals. These sub-techniques provide a more granular understanding of the techniques used by adversaries.
    
    \item Finally, documented adversary usage of techniques, procedures, and other metadata provides context around how adversaries use specific tactics and techniques. This metadata helps security professionals to identify patterns in adversary behaviour and develop more effective defence strategies.
\end{itemize}

By incorporating these components into its framework, MiTRE ATT\&CK enables security teams to gain a deeper understanding of how cyber adversaries operate. This understanding helps security professionals to develop proactive defence strategies that can identify and mitigate potential threats before they result in a security incident.

\section{Dataset}\label{dataset}

\begin{figure}[ht]
    \centering
    \includegraphics[scale=0.45]{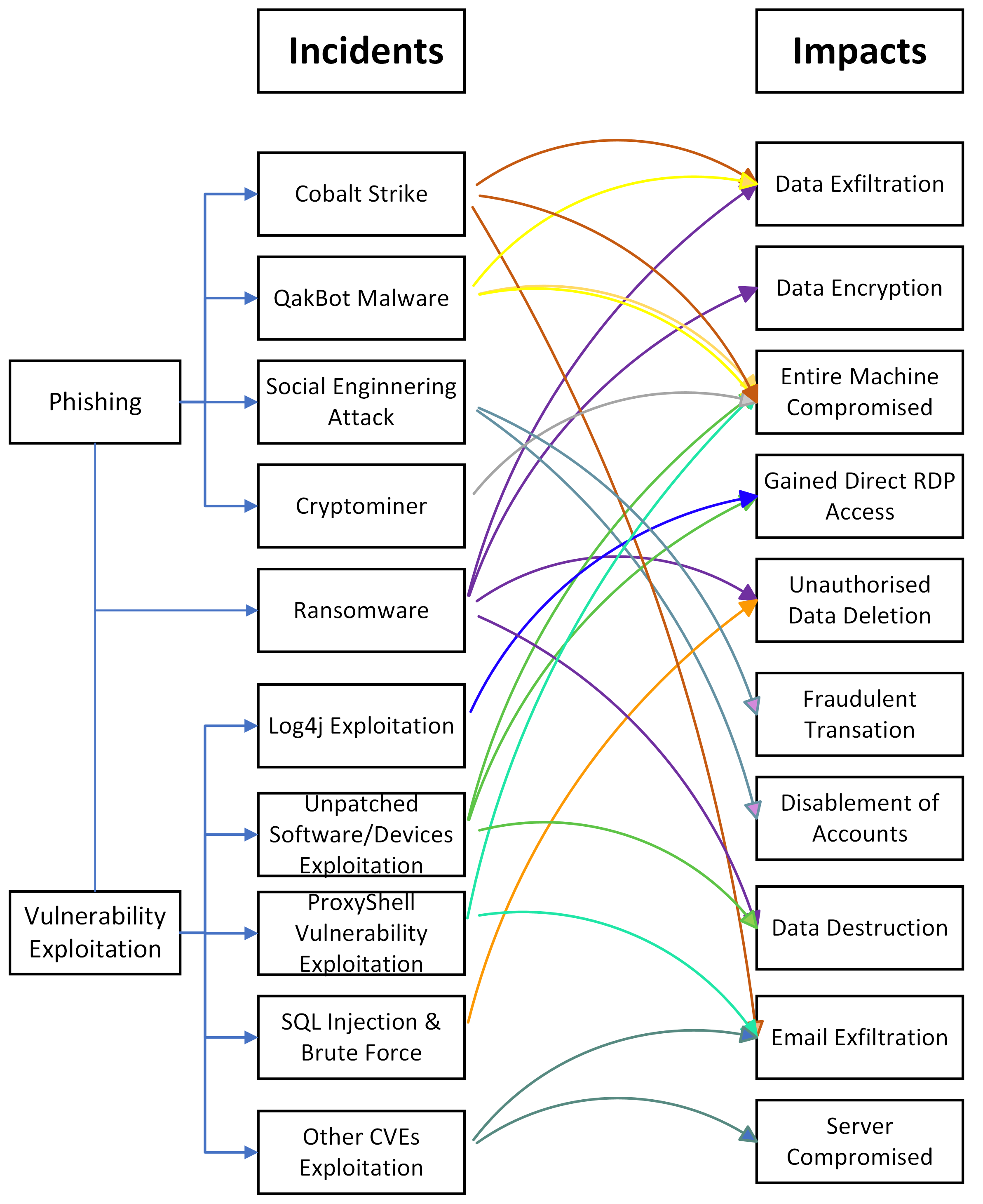}
    \caption{Type of Incidents and its Impacts}
    \label{fig:dataset}
\end{figure}

The data was provided by an organisation that delivers managed security services. They provided us with incident reports from real breaches; hence we keep their and their clients' identities confidential. We received a total of 50 real-life incidents, out of which we were able to use 45 for modelling; the remaining five (5) were not usable due to lack of detail. As depicted in Figure \ref{fig:dataset}, the majority of incidents were related to Phishing and \ac{VE}. Phishing incidents were further categorised into Cobalt Strike \& QakBot Malware Execution, Cryptominer, and Social Engineering Attacks. On the other hand, the \ac{VE} category included Unpatched Software/Device Exploitation, Log4j Exploitation, ProxyShell Vulnerability Exploitation, SQL Injection \& Brute Force, and Other CVEs Exploitation. Ransomware originated from both phishing and \ac{VE}. The severity of the incidents varied and caused different impacts as shown in Figure \ref{fig:dataset}, such as ransomware attacks caused data encryption, data exfiltration, unauthorised data deletion and data destruction. A different range of colours have been used to segregate the incidents that are associated with multiple impacts.

\section{Method}\label{method}

Our research methodology involved a multi-stage process that included the independent modelling of \acp{IFT} by the primary researcher, followed by review and feedback from a second researcher, and a subsequent collaborative discussion resulting in updates to the \ac{IFT} models.

During the first stage, the primary researcher worked independently to create an \ac{IFT} model for each incident under investigation. The \ac{IFT} model provides a graphical representation of the various events and factors that contributed to the incident, helping to identify its root causes and contributing factors.

In the second stage, the \ac{IFT} models were handed over to the second researcher for review. The second researcher carefully reviewed the \ac{IFT} models to identify any discrepancies or omissions and suggested necessary improvements (i.e., arguing to consensus~\cite{johnstone2017discourse}). In case of disagreements, they were referred to a third researcher to resolve them through discussion and brainstorming. This process ensured that the \ac{IFT} models were as comprehensive and accurate as possible.

\begin{figure*}[ht]
    \centering
    \captionsetup{justification=centering}
    \includegraphics[width=\textwidth]{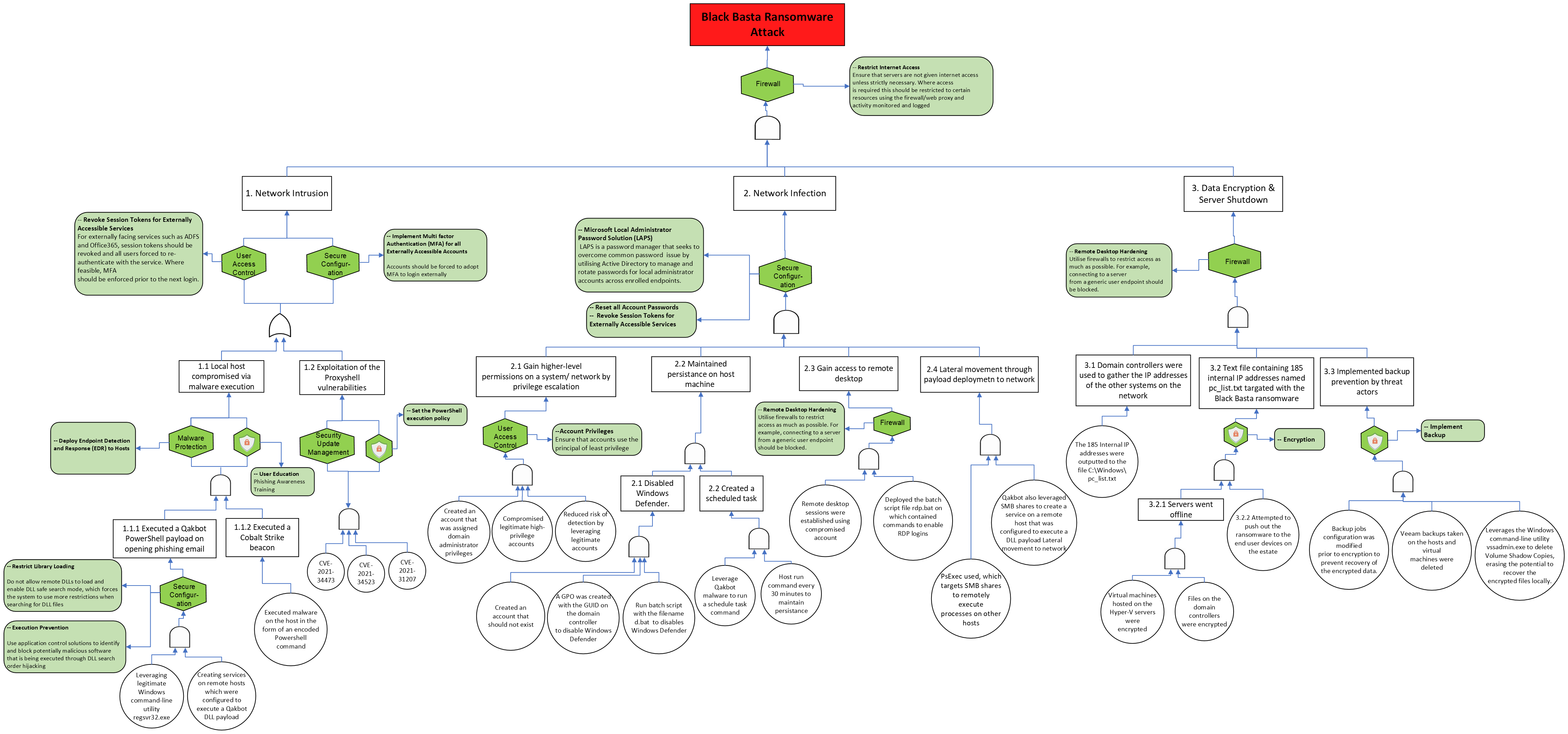}
    \caption{\ac{IFT} Model for Black Basta Ransomware Attack }
    \label{fig:case1}
\end{figure*}

\subsection{IFT Modelling}

The IFT modelling usually began by first reading the incident reports. The incident data were mapped to MiTRE ATT\&CK \ac{TTP}s. These reports include a timeline, executive summary, and remediation actions for each incident. We first started reading the executive summary to identify the type of incident and its impacts. The report began with initial access, which was usually the first stage in ATT\&CK. From there, the report detailed the steps taken by the attacker to carry out their attack.

Based on the information in the incident report, the first intermediate event was identified, which usually involved a network intrusion. This was followed by network infection and a third intermediate event based on the impact of the attack. The report was then studied in detail to identify each of the leaf nodes, which are the basic events that lead to the top intermediate events. 

We have shown one of the \ac{IFT} models from the dataset in Figure~\ref{fig:case1}. The incident is about the Black Basta ransomware attack that resulted in data exfiltration and server shutdown in an organisation. \ac{IFT}s help identify the confluence of sub-trees that facilitate an attack. In this case \emph{network intrusion}, \emph{network infection} and \emph{data encryption \& server shutdown} sub-trees contribute to the success of the attack. In our diagram we denote them as 1, 2 and 3 respectively. If we drill down into each of them we identify the \emph{basic events}, \emph{intermediate events} and \emph{inhibit gates} that form the individual sub-trees. They are numbered accordingly; for example, an \emph{intermediate event} for 1 will be numbered as 1.x, so on and so forth for each of the other sub-trees. 

For \emph{network intrusion}, the attacker exploited legitimate Windows command line utility and the ability to execute services on a remote host. We classify them as \emph{basic events} as they are the core enablers and form the leaf nodes for \emph{network intrusion}. The inhibit gate identifies the mitigation for the leaf nodes in \emph{network intrusion}, which is a secure configuration. The absence of secure configuration allows the execution of Qakbot PowerShell payload and Cobalt Strike beacon as 1.1.1 and 1.1.2 of Figure \ref{fig:case1}. While the \ac{IFT} reconstructs the path, it also identifies the intersection where mitigation could have helped and, in turn, the Cyber Essential control and/or \ac{AC}. The separation between the enabler and the consequent action enables a nuanced understanding of incidents and corresponding mitigations. The other \emph{inhibit gates} in this sub-tree, such as malware protection and security update management, protect against malware and PowerShell vulnerabilities, respectively. There is an additional control of user education which can potentially prevent users from clicking links in suspicious emails.  The incident report does not conclusively note the cause of the \emph{network intrusion} and points to either relevant CVEs related to PowerShell or malware downloaded through a phishing email. We represent this using an \emph{OR gate}. Finally, \emph{network intrusion} could have been prevented through user access control and/or secure configuration.   

Similarly, for \emph{network infection}, creation of accounts, disabling windows defender and scheduling a task form the basic enablers and are the \emph{basic events}. The \emph{inhibit gate} points to access control which could have prevented the malware from creating user accounts bypassing the privilege management infrastructure. A firewall is identified as a control at the \emph{inhibit gate} which could have prevented inbound connection to a remote desktop along with remote desktop hardening. PsExec and Qakbot exploited SMB shares to enable lateral movement. Privilege escalation, persistence, access to a remote desktop and lateral movement led to \emph{network infection}. This attack sub-tree could have been prevented through secure configuration.

Finally, \emph{data encryption \& server shutdown} were enabled by encrypting the virtual machines hosted on Hyper-V servers and files on domain controllers. They are classified as \emph{basic events} along with the attacker's ability to push out ransomware on end user devices in the estate. The intermediate events, such as gathering IP addresses, targeting internal IP addresses and preventing backups led to \emph{data encryption \& server shutdown}. These actions could have been prevented with an appropriately configured firewall since the attacker executed the attack remotely.

% The intermediate events, such as gathering IP addresses, targeting internal IP addresses and preventing backups led to \emph{data encryption \& server shutdown}. Since they were executed remotely by the attacker so an appropriately configured firewall could have prevented this particular attack sub-tree. \hl{Alternative???}\textcolor{blue}{The attacker executed intermediate events, such as gathering IP addresses and preventing backups, which ultimately led to\emph{data encryption \& server shutdown}. These actions could have been prevented with an appropriately configured firewall since the attacker executed the attack remotely.}

\subsection{Identifying Security Controls}
In the \ac{IFT}s, we employed three different kinds of controls: \ac{CE} Controls, controls provided in the incident reports and controls we identified as relevant. After the completion of the model, we initiated an analysis of the potential intermediate events or basic events that \ac{CE} can prevent. % Subsequently, we set up inhibit gate and formulated a list of recommendations that included in Cyber Essentials. 
The security measures described in the incident report were also used for this purpose. Several of the suggested controls from the reports were already present in \ac{CE}. We then proceeded to look for events that needed  \ac{AC} (either recommended in the reports or ones we identified from the literature). Once we identified controls for all the inhibit gates, two researchers discussed the best way to finalise the controls and concluded the process. As illustrated in Figure \ref{fig:case1}, inhibit gates with text inside are Cyber Essential controls and inhibit gates with shield symbol inside are \ac{AC}.

\section{Analysis}\label{analysis}

\paragraph{\textbf{Analysis by Edges}}

\begin{figure}[ht]
    \centering
    \includegraphics[scale=0.40]{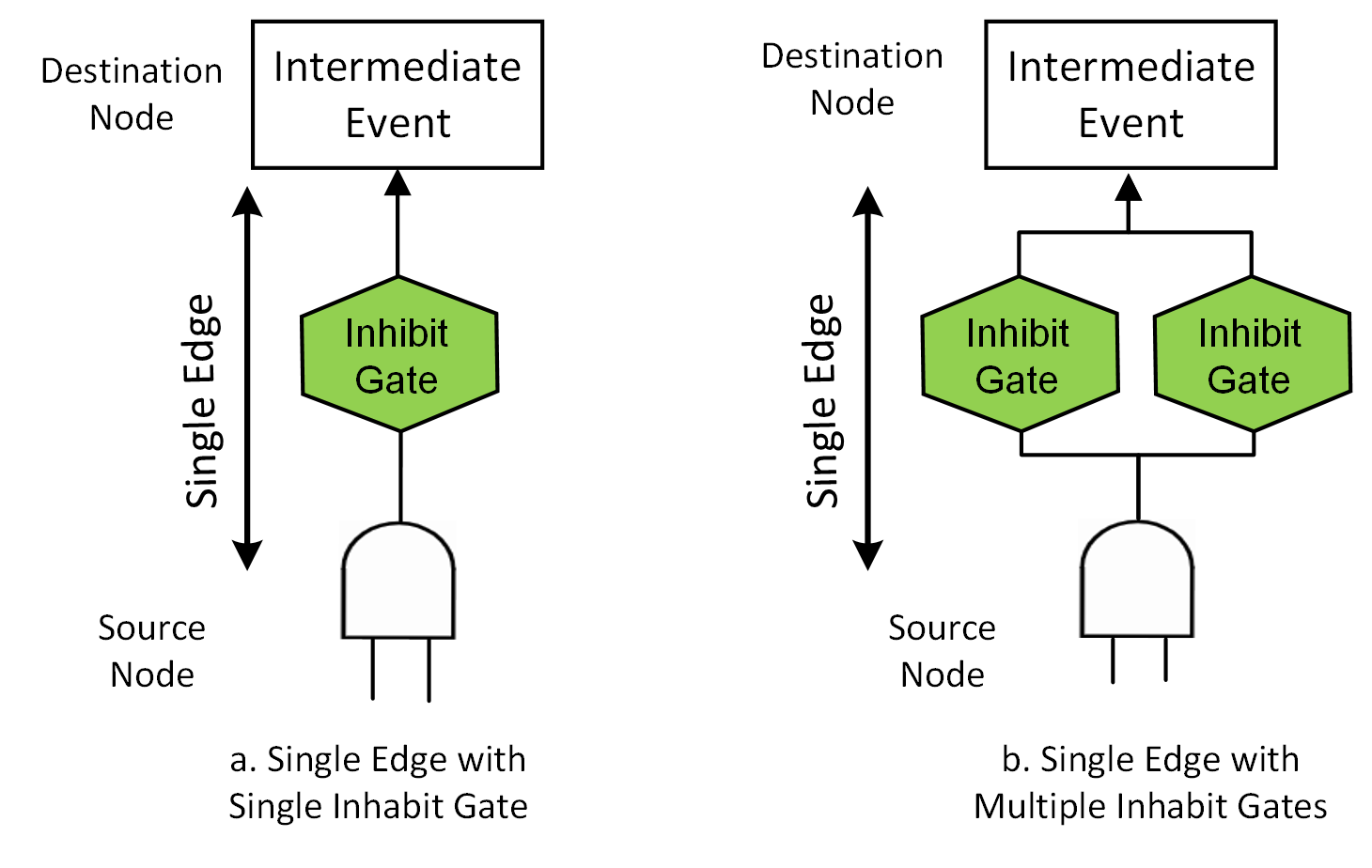}
    \caption{Single Edge between And Gate and Intermediate Event}
    \label{fig:edge}
\end{figure}

To determine the effectiveness of \ac{CE}, we counted the number of edges that could have been prevented by \ac{CE} in each \ac{IFT}. Our definition of \emph{edge} involves linking two nodes by an inhibit gate. So we consider each instance of the inhibit gate in the \ac{IFT} model as a single edge as shown in Figure \ref{fig:edge} (a). If there are several inhibit gates connecting the source and destination node, we also treat them as a single edge (Figure \ref{fig:edge} (b)). In every case, the source node is an AND/OR gate, and the destination node is an intermediate event.

\paragraph{\textbf{Analysis by Level}}

\begin{figure}[ht]
    \centering
    \includegraphics [width=5cm, height=5cm]{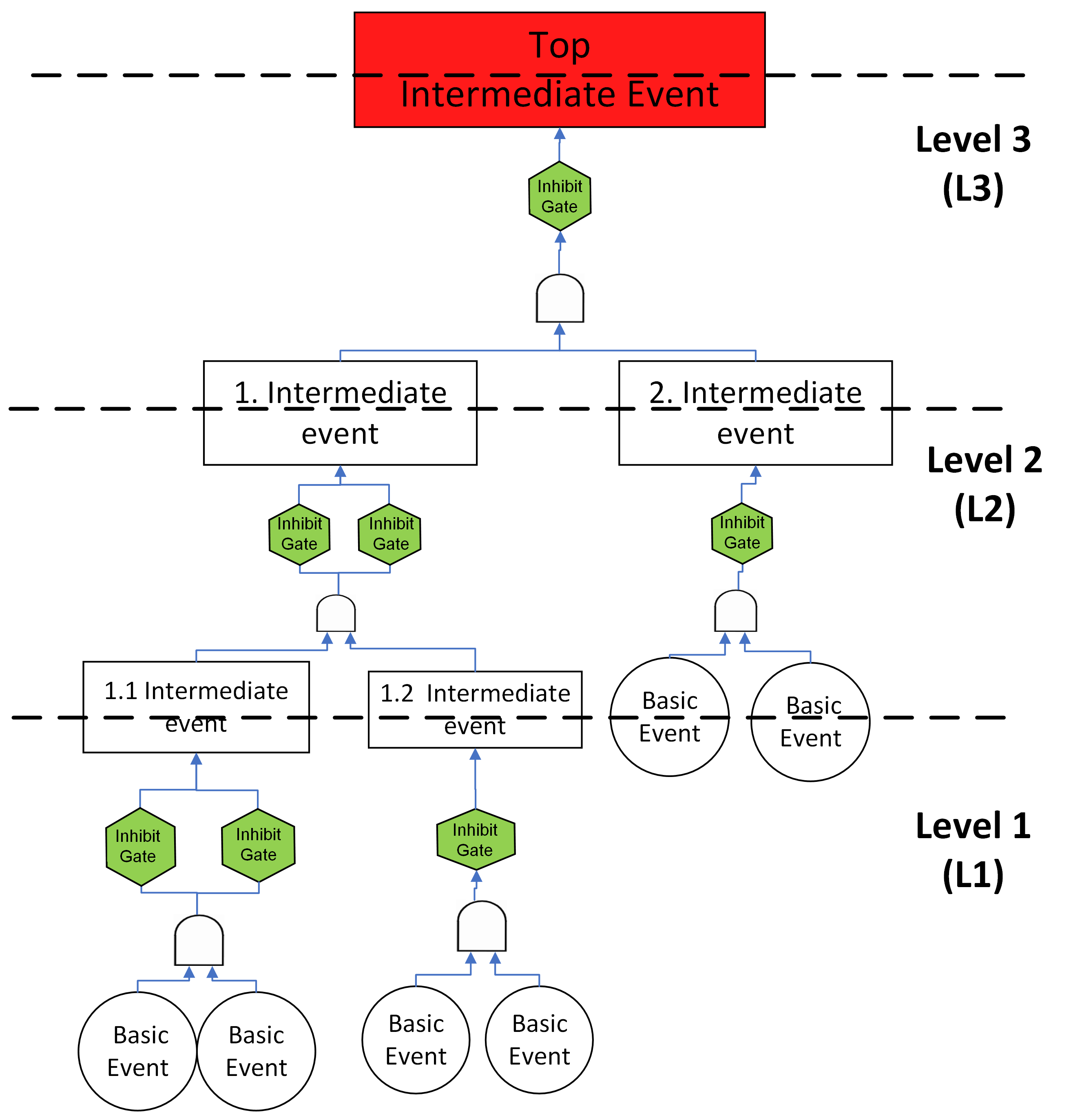}
    \caption{Three Level \ac{IFT}}
    \label{fig:level}
\end{figure}

For this type of analysis, we first examine the inhibit gates at the lowest \emph{level} within each tree. As illustrated in Figure~\ref{fig:level}, a three-level tree structure is depicted, with levels denoted as L1, L2 and L3. The first two layers, L1 and L2, have two edges and three inhibit gates each. The top layer, L3, has only one edge and one inhibit gate. We focus on L1 of each tree and identify which paths are restricted by \ac{CE}. By examining the lowest level, we can determine the effectiveness of the \ac{CE} controls in preventing an attack from going further in the tree. If the \ac{CE} controls prove inadequate, we then proceed to examine \ac{AC} controls, as well as potential combinations of both.

\paragraph{\textbf{Analysis by Phase}}

\begin{figure}[ht]
    \centering
    \includegraphics[width=5cm, height=4.5cm]{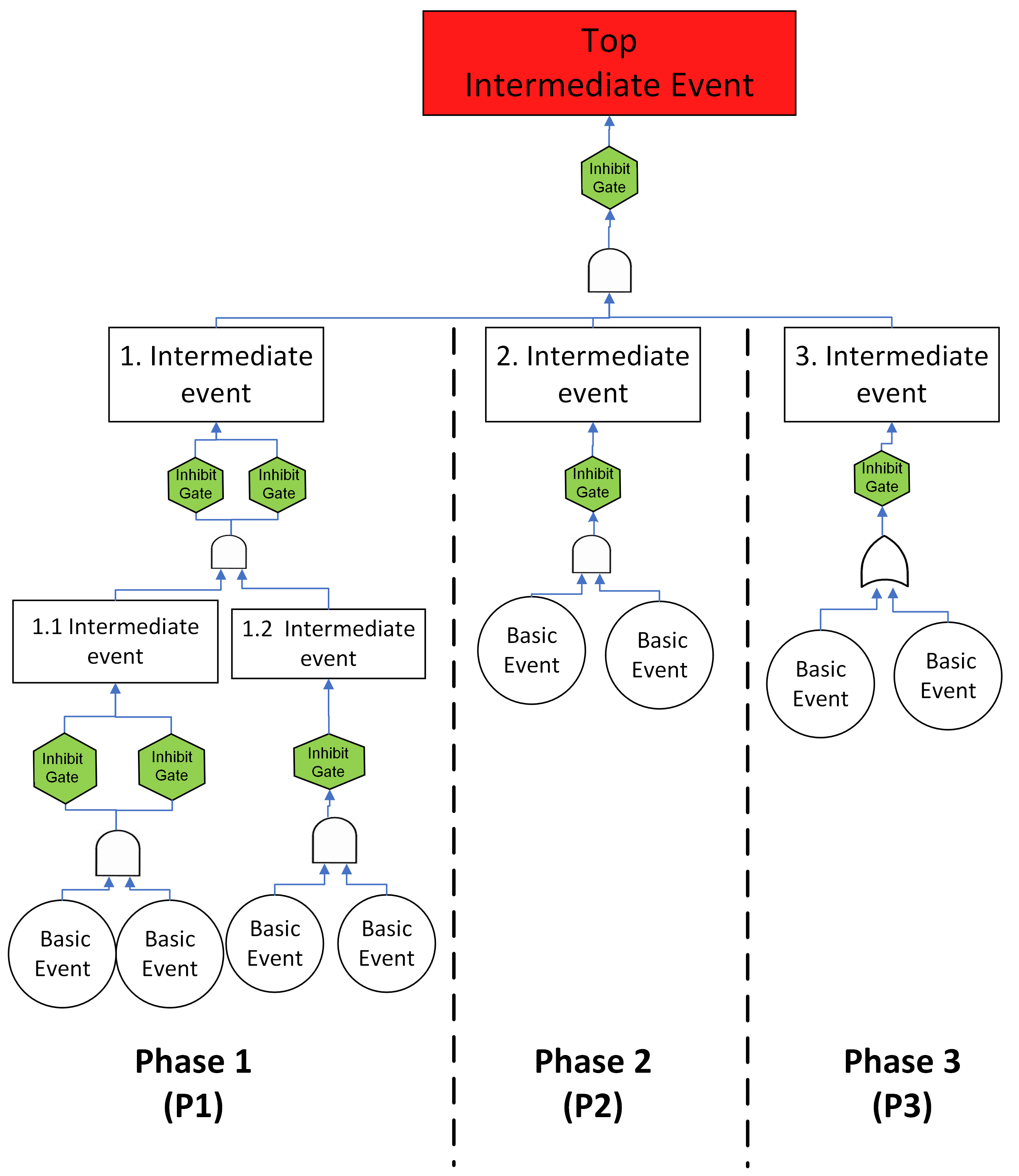}
    \caption{Three Phase \ac{IFT}}
    \label{fig:phase}
\end{figure}

In the phase-wise analysis, we divide the \ac{IFT} model into distinct phases representing intermediate events occurring after the top-level event. Figure \ref{fig:phase} illustrates a three-phase structure consisting of Phase 1 (P1), Phase 2 (P2), and Phase 3 (P3). P1 is the initial phase, the first intermediate event following the top level event. We analyse attack mitigation by \ac{CE} in its initial phase P1. This allows us to determine the level of protection provided by \ac{CE} in preventing attacks during the initial phase. Additionally, we examine the use of other controls, such as \ac{AC} and combinations of \ac{CE} and \ac{AC} that are employed in P1.

\paragraph{\textbf{Analysis by Level+Phase}}

\begin{figure}[ht]
    \centering
    \includegraphics[width=5cm, height=4.5cm]{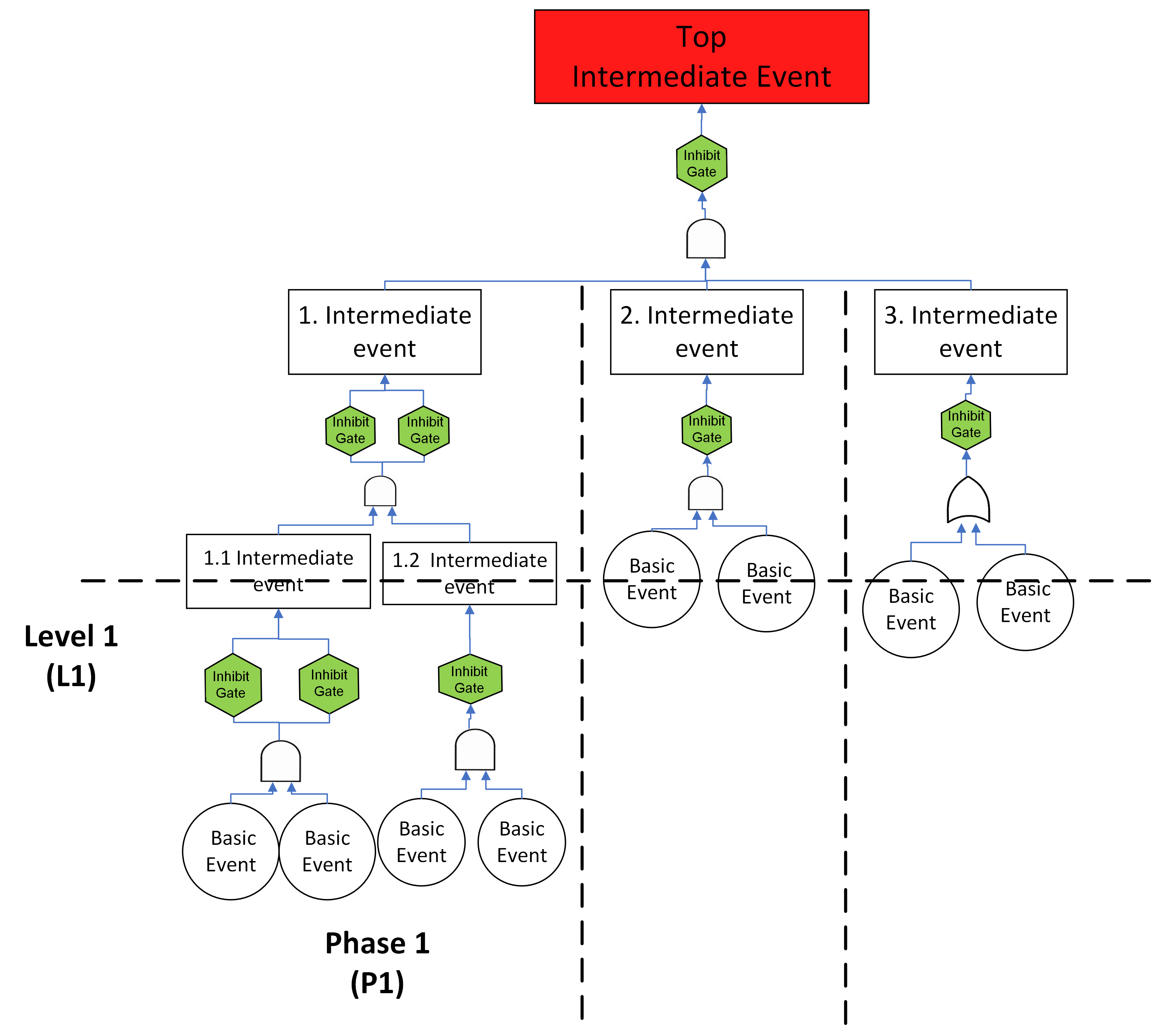}
    \caption{Level 1 (L1) + Phase 1 (P1) Analysis}
    \label{fig:l1p1}
\end{figure}

In this analysis, we integrated both level-wise and phase-wise analysis as illustrated in Figure \ref{fig:l1p1}. To begin with, we focused on examining the attack paths at the lowest level (L1) during the initial phase (P1), which we referred to as L1P1. Subsequently we analysed the security controls implemented at the corresponding L1P1. By combining these two insights, we delved deeper into assessing the effectiveness of both \ac{CE} and \ac{AC} controls. This comprehensive analysis allowed us to gain a more thorough understanding of the overall effectiveness of these controls to stop an attack during the initial stages.

\section{Findings}\label{findings}

\subsection{RQ1: Effectiveness of \ac{CE} in inhibiting attack pathways.} 

The data presented in Figure \ref{fig:bar1} indicates the attacks inhibited by \ac{CE} controls. Firewall and Secure Configuration were applicable in 39 out of 45 incidents, making them the most effective security measures in blocking malicious traffic and preventing unauthorised access. This is a promising indication that the attack pathways can be inhibited by proper configuration to minimise the risk of cyber attacks. 
User Access Control was suitable in 22 out of 45 incidents. Malware Protection was applicable in 25 out of the 45 incidents. Lastly, Security Update Management, which involves regularly updating systems and software to address security vulnerabilities, was only applicable in 19 out of the 45 incidents. 

\tikzset{font=\footnotesize}
\begin{figure}[htb]
\begin{tikzpicture}
\begin{axis}[ legend style={at={(0.5,-0.1)},anchor=north, legend columns=2},
    xbar,
    xmin=0,
    symbolic y coords={{Firewall}, {Secure Configuration}, {User Access Control}, {Malware Protection}, {Security Update}},
    ytick=data,
    nodes near coords, 
    nodes near coords align={horizontal},
    enlarge y limits=.3,
    height=5.5cm, width=7cm,
]
\addplot[fill=blue!30,draw=blue!90,tickwidth = 0pt,bar width=5pt,, tick label style={font=\small}] coordinates {
(39,{Firewall}) 
    (39,{Secure Configuration}) 
    (22,{User Access Control}) 
    (25,{Malware Protection})
    (19,Security Update)
    };
\end{axis}
\end{tikzpicture}
\caption{Inhibitors of attacks by CE controls}
    \label{fig:bar1}
\end{figure}
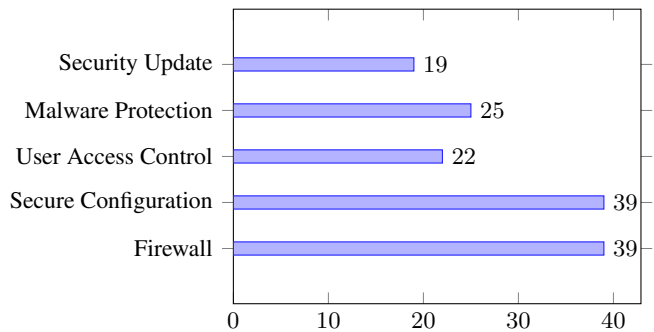

\begin{table*}
\tiny
% \begin{table}[h!]
\centering
 \begin{tabular}{p{2.5cm} |p{0.6cm} |p{0.6cm} |p{0.6cm}|p{0.8cm}|p{0.5cm}|p{0.5cm}|p{0.8cm}|p{0.5cm}|p{0.5cm}|p{0.8cm}|p{0.5cm}|p{0.5cm}|p{0.8cm}} 
 \hline
 Cases & Total Edges  & CE Edges& AC Edges & CE+AC Edges & CE at L1& AC at L1 & CE+AC at L1 & CE at P1 & AC at P1 & CE+AC at P1 & CE at L1P1 &AC at L1P1 & CE+AC at L1P1\\ [0.5ex] 
 \hline
01(Ransomware) & 11 & 6 & 3& 2 &  1 & 2  & 0 & 2 &   0   &  2   &  1 & 0  & 0  \\ \hline
02(Phishing) & 5 & 1 & 3& 1 &  0 & 1  &0 &  0 &     1 &     1 &  0 &  1 &  0   \\\hline
03(Malware Execution) & 7 & 3 & 3 & 1 &  0 & 1  &1 &1  &   1   &    1  &  0 & 1  &  1   \\\hline
04(Malware Execution) & 11 & 8 & 1 & 2 &  1 & 0  &1 &1 &    0  &    1  & 1  & 0  &   0  \\\hline
05(Malware Execution) & 8 & 5& 1 &2 &  4 & 0  &0 & 1  &      0&    1  &  1 &  0 &   0  \\ \hline 
06(Phishing) & 4 &2  &1  &1 & 1  & 0  & 0 &0 &  0    &   1  &  0 &  0 &   1   \\ \hline
07(Ransomware) & 10& 4 & 4 & 2 & 0&1&0 & 2     &   1   &   0  &  1 &  1 &    0 \\ \hline
08(Phishing) & 3 & 0 & 2 & 1 &  0 & 2  &1 &0   &   0   &  1  &  0 &  0 &   1     \\\hline 
09(Malware Execution) & 4 & 2 & 0 &  2&  2 &0   &1 &1  &    0  &    0  & 1  & 0  &  0    \\\hline  
10(Malware Execution) & 4  & 4 & 0 & 0& 4 &0   &0 & 1  &     0 &   0  & 1  &  0 &  0     \\\hline 
11(CV Exploitation) & 4 & 3 &1  & 0 & 1  & 0  &0 &2  & 0     &   0  & 1  & 0  &  0     \\ \hline 
12(CV Exploitation) & 3 & 2 &1  &0 & 1  & 1  &0 &1  &   0   &    0  & 1  &  0 &  0     \\\hline 
13(CV Exploitation) & 2 & 1 & 1 & 0 &   1& 1  &0 &1  &   0   &   0   &1   &  0 &  0    \\\hline 
14(Phishing) & 4 & 2  & 1 & 1 &  1 & 0  &1 &0  &   0   &   1  &  0 &  0 &   1   \\\hline 
15(Phishing) & 2 &  0& 0 & 2 &  0 & 0  &2&0   &     0 &    1  &  0 &  0 &   1   \\\hline
16(CV Exploitation) & 3 & 2 & 0 & 1 &  1 & 0  &1& 1  &0      &    0  & 1  &  0 &   0   \\ \hline 
17(CV Exploitation) & 4 &2  &1  & 1 &  1 & 1  &1&1  &  0    &    0  &  1 & 0  &    0   \\\hline
18(CV Exploitation) & 4 & 1 & 1 & 2 &   0& 1  &0&0   &  0    &   1  & 0  &  0 &    1   \\\hline 
19(CV Exploitation) & 2 & 1 & 1 & 0 &   1&  1 & 0 &1  &  0    &  0  & 1  &  0 &    0   \\\hline
20(Phishing) & 3 & 1 & 0&2 &   0& 0  &2 &  0 &    0  &    1  &  0 & 0  &    1   \\\hline
21(CV Exploitation) &3& 2 & 0 & 1 &  1 & 0  &1  & 1  &   0   &   0  &  1 & 0  &    0   \\ \hline
22(Phishing) & 3 & 2&0&1 &  1 & 0  & 1   & 0 &    0  &    1  & 0  & 0  &     1  \\ \hline
23(Malware Execution) & 3 & 1 & 0 & 2& 1 &   0& 1 & 0  &  0    &   1  &  0 & 0  &     1  \\  \hline
24(Phishing) & 2 &0 & 0 & 1 &  0 & 0  & 1&0  &   0   &    1 & 0  &  0 &      1  \\ \hline
25(Ransomware) & 6 &2 &1 & 3  &  1 & 1  &0 &1  &   0   &   1  & 1  & 0  &      0  \\ \hline
26(Phishing) & 3 &2 &0 &1 & 1  &  0 & 1 & 0  &  0    &   1  & 0  & 0  &      1  \\\hline
27(CV Exploitation) & 3 &2 &1 & 0 &  1& 1  &0 &  1 &   0   &    0  & 1  &  0 &    0    \\ \hline
28(Ransomware) & 7 & 3& 2& 2&   2& 1  &   0 &1 &  0    &    1 & 1  & 0  &    0    \\ \hline
29(CV Exploitation) & 3 & 2& 0& 1 &   1& 0  & 1 &1  &   0   &    0  & 1  & 0  &  0     \\ \hline
30(Malware Execution) & 4 & 1 &1 & 2 &   0&   1&   0 &0 &   0   &  1  &  0 &  0 &    1    \\ \hline
31(Ransomware) & 6 & 3& 1&  2 &1   & 1  &0 &2   &    0  &    0  & 1  & 0  &  0    \\ \hline
32(Phishing) & 3 & 3 & 0 & 0 &  2 &   0&   0&1  &  0    &   0  & 1  & 0  & 0     \\ \hline
33(Ransomware) & 3 & 2 & 0 & 1  & 1  & 0  & 1  &0 &   0   &    1 & 0  &  0 &  1   \\ \hline
34(CV Exploitation) & 3 & 2 & 0 & 1  & 1  &0   &  1  &1 &  0    &   0  & 1  &  0 &  0  \\ \hline
35(Phishing) & 3 & 1 & 0 & 2  &  0 & 0  & 2  & 0 &  0    &   1 &  0 &  0 &  1   \\ \hline
36(Malware Execution) & 6 & 4 & 2 & 0  & 1 & 1  & 0  &2 &   0   &     0 & 1  & 0  &  0   \\ \hline
37(CV Exploitation) & 6 & 3 & 3 & 0 & 0  & 2  & 0& 1   &  0    &    0 &  1 &  0 & 0    \\ \hline
38(Ransomware) & 8 & 3 & 2 & 3 &   2&  2& 0 & 1  &   0   &    1 & 1  &   0&  0    \\ \hline
39(Ransomware) & 6 & 3 & 2 & 1 &  1 &  2 &  0 &1 &   0   &    0  & 1  & 0  &  0   \\ \hline
40(Ransomware) & 9 & 5 & 2 & 2  & 3  &2  &  1 & 2 &  0    &   1  &  1 &  0 &  1   \\ \hline
41(CV Exploitation) & 3 & 2 & 0 & 1  & 1 & 0  &  1 & 1 &  0   &     0 &  1 &  0 &  0  \\ \hline
42(Phishing) & 5 & 2 & 2 & 1  &  1 & 1  &  0  &0 &  0    &   1  &  0 & 0  & 1   \\ \hline
43(Phishing) & 5 & 2 & 2 & 1  &  1 & 1  & 0  &  0 &  0    &   1 & 0  &  0 & 1   \\ \hline
44(Phishing) & 4 & 3 & 0 & 1  &  1 & 0  & 0  &  0 &  0    &  1  &  0 &  0 &  1  \\ \hline
45(Phishing) & 3 & 2 & 0 & 1  &  1 & 0  & 1 & 0  &   0   &    1  &  0 &  0 & 1  \\ [1ex]
 \hline
\end{tabular}
 \caption{Attack Pathways Stopped by Cyber Essentials (CE), Additional Controls (AC) and Combination of Both (CE+AC)}
    \label{table:table1}
\end{table*}

We present our findings in Table \ref{table:table1}. As noted above, we took a holistic approach to answering this research question by evaluating each component of the attack tree under four distinct analytical outcomes: analysis-by-edge, analysis-by-level, analysis-by-phase and analysis-by-level+phase. 

\paragraph {\textbf{Analysis by Edge}} The mitigation of attack pathways in the analysis-by-edge approach are represented in the three columns of the Table~\ref{table:table1} starting from third column, i.e., CE Edges. The Case 01 in Table~\ref{table:table1} has 11 edges, of which 6 edges are inhibited by \ac{CE} controls, 3 through \ac{AC} and 2 edges are inhibited by a combination of both \ac{CE} and \ac{AC}. Case 04 too has 11 edges, where 8 edges are mitigated through \ac{CE} controls, 1 edge through \ac{AC} and 2 edges through a combination of both \ac{CE} and \ac{AC}. These two incidents have the maximum coverage of \ac{CE} controls, yet \ac{AC} are required for comprehensive mitigation of the threats.

There are 18 cases where only 2 edges are inhibited by \ac{CE} controls alone. However, the total number of edges for these  18 cases ranges from 3 to 6. This means there remains the need for at least one \ac{AC} along with the relevant \ac{CE} control to mitigate each of the 18 incidents.  

Case 07 has 10 edges, where 4 edges can be prevented through \ac{AC} - which is the highest among all the cases requiring \ac{AC}. There are 17 cases where a combination of \ac{CE} and \ac{AC} controls mitigate threats, but these cases do not have independent \ac{AC} edges. So \ac{AC} on their own cannot be effective without the relevant \ac{CE} controls. Only two cases, namely 25 and 38, have the highest combinations of \ac{CE} and \ac{AC} edges, amounting to 3. Again, this shows that neither \ac{CE} nor \ac{AC} are effective on their own. There are 20 cases which have only 1 combined edge each. This indicates that a significant proportion of our cases require some combination of \ac{CE} and \ac{AC} (in this case, at least 1) to prevent attacks. % When these cases are looked in conjunction it indicates a reasonable level of requirement of additional controls.

\paragraph {\textbf{Analysis by Level} }The analysis-by-level also highlights the need for a combination of \ac{CE} and \ac{AC}. \ac{CE} controls can only mitigate 4 edges from  2 of the 45 incidents at level 1. \ac{AC} can prevent at most 2 edges at level 1, and they appear in 6 cases. A combination of both \ac{AC} and \ac{CE} prevent 2 edges at level 1 from 2 of 45 cases. % Level 1 is primarily made up of enablers or \emph{basic events} which are inherent properties/configurations of the underlying platform. % Our investigations reveal that erroneous user interactions such as clicking phishing emails helps negatively exploit such enablers. 

% Overall, our investigations reveal that \ac{CE} is moderately effective and there is a need for \ac{AC}. 
\paragraph {\textbf{Analysis by Phase} }During the phase-wise analysis, we focused on Phase 1 (P1) and examined the edges of the IFT that are mitigated by \ac{CE}, \ac{AC} and CE+AC. The analysis of this method is depicted in the Table \ref{table:table1} column named \emph{CE at P1}, \emph{AC at P1} and \emph{CE+AC at P1}. When we examined the attack paths mitigated by use of \ac{CE} in P1, our analysis revealed that there were 6 cases where 2 edges were inhibited, 21 cases where 1 edge was inhibited, and 18 cases where no edges were inhibited. Regarding the use of \ac{AC} in P1, there were 42 cases where no \ac{AC} was required and 3 cases where 1 edge was inhibited by \ac{AC}. The combination of controls in P1 showed that there were 19 cases where no edges were inhibited, 25 cases where 1 edge was inhibited, and 1 case where both controls stopped 2 edges.

The analysis highlights that \ac{CE} played a crucial role in stopping attacks during the initial stage in most cases. However, layering with \ac{AC} in some instances could further enhance the security posture by inhibiting attack pathways.

\paragraph {\textbf{Analysis by L1P1}} The combination of level-wise and phase-wise analysis revealed that \ac{CE} plays a major role in stopping the malicious edges. The findings are presented in the last three columns in Table \ref{table:table1}. There are no more than two edges appearing in the L1P1 of IFT. The majority of 1's are occupied by \ac{CE} alone, and there are only three cases - 02, 03 and 07 - where \ac{AC} alone are used. This shows that \ac{CE} alone is very effective in stopping the threats at an early stage. However, the number of edges in CE+AC should also be taken into account, and a total of 19 cases are observed to be mitigated by combined controls. This analysis shows that even though \ac{CE} plays a major role as a standalone control, a combined approach provides further defence.

\begin{table}[ht]
\small
  \centering
  \begin{tblr}{
    colspec = {|p{3cm}|p{2cm}|p{2cm}|p{2cm}|p{2cm}|},
    % row{2} = {blue7},
    % row{3} = {purple7},
    % row{4} = {teal7},
    % row{5} = {yellow7},
    % column{3} = {teal7},
    % cell{2}{3} = {yellow7},
  }
    \hline
    \textbf{Method}& \textbf{Total attack pathways} & \textbf{Inhibited by CE} & \textbf{Inhibited by AC} & \textbf{Inhibited by CE+AC}  \\\hline
    \hline
        Edge Analysis &  209 &  108 &  46 & 55  \\
    \hline
        Level Analysis &  98 &  46 &  28 & 24  \\
    \hline
    \hline
    \textbf{Method }& \textbf{Total Cases} & \textbf{Mitigated by CE} & \textbf{Mitigated by AC} & \textbf{Mitigated by CE+AC}  \\\hline

        Phase Analysis &  45 & 17 &  0 & 28  \\
    \hline
        Level+Phase Analysis &  45 &  25 &  1 & 19 \\
    \hline

    % \hline
    %     Phase Analysis &  63 & 33 &  3 & 27  \\
    % \hline
    %     Level+Phase Analysis &  48 &  26 &  3 & 19 \\

    \end{tblr}
  \caption{Overall Summary of attacks inhibited by CE \& AC controls in each analysis approach}
    \label{table:table2}
\end{table}

The summarised findings are presented in Table \ref{table:table2}, encapsulating the key insights from the analysis. It is evident that in the Edge and Level analysis, the majority of attack pathways are effectively inhibited by the implementation of \ac{CE} controls alone. However, it should be noted that \ac{AC} measures also play a vital role in completely halting the attacks. In the Phase analysis, 17 out of 45 cases were successfully stopped solely by \ac{CE} controls, while 28 cases required a layered approach incorporating both \ac{CE} and \ac{AC} controls. Particularly noteworthy is the effectiveness of \ac{CE} controls in the L1P1 analysis, where 25 out of the 45 attacks were exclusively prevented by \ac{CE} measures. This analysis underscores the significance of deploying \ac{CE} controls at the lower levels of attacks during their initial phases, as it can effectively impede the spread of attacks further. 

\subsection{RQ2: Which \ac{AC} appear frequently and how effective are they as inhibitors of attack pathways.}

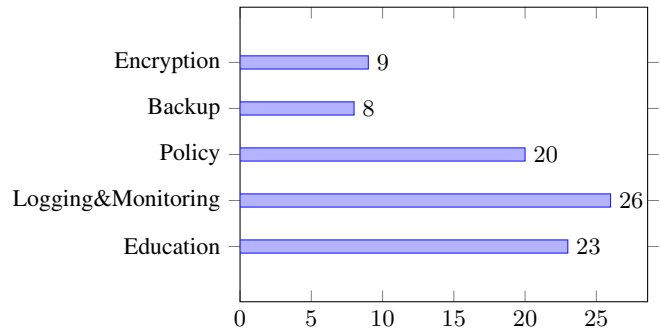
\begin{figure}[htb]
\small
\centering
\begin{tikzpicture}
\begin{axis}[ legend style={at={(0.5,-0.1)},anchor=north, legend columns=2},
    xbar,
    xmin=0,
    symbolic y coords={{Education},{Logging\&Monitoring}, {Policy},{Backup},{Encryption}},
    ytick=data,
    nodes near coords, 
    nodes near coords align={horizontal},
    enlarge y limits=.3,
    height=5.5cm, width=7cm,
]
\addplot[fill=blue!30,draw=blue!90,tickwidth = 0pt,bar width=5pt,, tick label style={font=\small}] coordinates {
(23,{Education}) 
(26,{Logging\&Monitoring})
    (20,{Policy}) 
    (8,{Backup})
    (9,{Encryption}) 
    };
%     \addplot[fill=red!30,draw=red!90,tickwidth = 0pt,bar width=5pt,, tick label style={font=\small}] coordinates {
% (10,{Education}) 
%     (70,{Logging\&Monitoring})
%     (10,{Policy}) 
%     (10,{Backup})
%     (10,{Encryption}) 
%     };
     % \legend{Phishing, Vulnerability Exploitation}
\end{axis}
\end{tikzpicture}
\caption{Inhibitors of attacks by Additional Controls}
    \label{fig:bar2}
\end{figure}

After conducting a thorough analysis of the data, we have identified five \ac{AC}, on top of \ac{CE}, that are frequently used to prevent attacks from succeeding. These controls include Encryption, Backup, Policy, Education, and Logging \& Monitoring. 
\begin{itemize}
    \item\textbf{Education:} This control refers to education and training for users on cyber security best practices. By providing employees with regular cyber security training, organisations can reduce the risk of human error and strengthen their overall security posture. From the analysis of the 45 cyber incidents, several pathways were successfully inhibited by implementing \ac{AC}. As illustrated in Figure \ref{fig:bar2}, education was effective in 23 out of 45 incidents. This suggests that educating employees about cyber security best practices and potential threats is crucial in reducing the likelihood of successful cyber attacks, such as phishing. 
    
    \item \textbf{Logging \& Monitoring:} Logging \& Monitoring involves collecting and analysing network activities and security events. These activities include monitoring network traffic, log files and end-user activities. This control is important for identifying security incidents and minimising the damage caused. Logging \& Monitoring was found to be effective in 26 out of 45 incidents in inhibiting attack edges. This indicates that closely monitoring system logs and network activities can effectively detect and prevent cyber attacks. 
    
    \item \textbf{Policy:} Cyber security policies are rules and guidelines that outline an organisation's approach to cyber security. By establishing and implementing clear policies and procedures, organisations can ensure that their security measures are consistently applied across the organisation. Configuring an appropriate execution policy was another control that proved effective in inhibiting attack pathways, being utilised in 20 out of the 45 incidents. 
    
    \item \textbf{Encryption:} Encryption involves encoding data so that it can only be read by authorised parties. The implementation of encryption can help the organisation to secure sensitive data and prevent unauthorised access. Encryption was used in 9 out of 45 incidents, highlighting the effectiveness of strong encryption measures to prevent unauthorised access to sensitive data.
    
    \item \textbf{Backup:} Backup refers to making regular replicas of data and storing them in a secure location. This is a crucial security measure in the case of ransomware, or other incidents where data may be damaged or deleted. By establishing a backup(s), organisations can ensure that data is accessible even in the case of disaster. We reconstructed 9 cases of ransomware, where backup, as a control, can support organisations to recover from such cyber attacks and minimise the impact of data loss. 
\end{itemize}

Overall, these findings suggest that implementing \ac{AC} with \ac{CE} can be a practical approach to reducing the attack surface and consequently the blast radius. 

% \tikzset{font=\footnotesize}

% \begin{table}[h!]
% \centering
%  \begin{tabular}{||p{1.5cm} |p{1cm} |p{1.26cm} |p{1.26cm}|p{1.3cm}||} 
%  \hline
%  Incidents & Education & Policy & Backup & Logging \& Monitoring \\ [0.5ex] 
%  \hline\hline
%  Phishing & 70\% & 25\% & 10\% & 60\%  \\ \hline
%  Vulnerability Exploitation & 10\%& 10\%  & 10\%& 80\% \\ [1ex] 
%  \hline
%  \end{tabular}                                 
% \end{table}

\subsection{RQ3: Identify specific inhibition patterns (\ac{CE} or \ac{AC}) with regards to ransomware incidents in the corpus.}

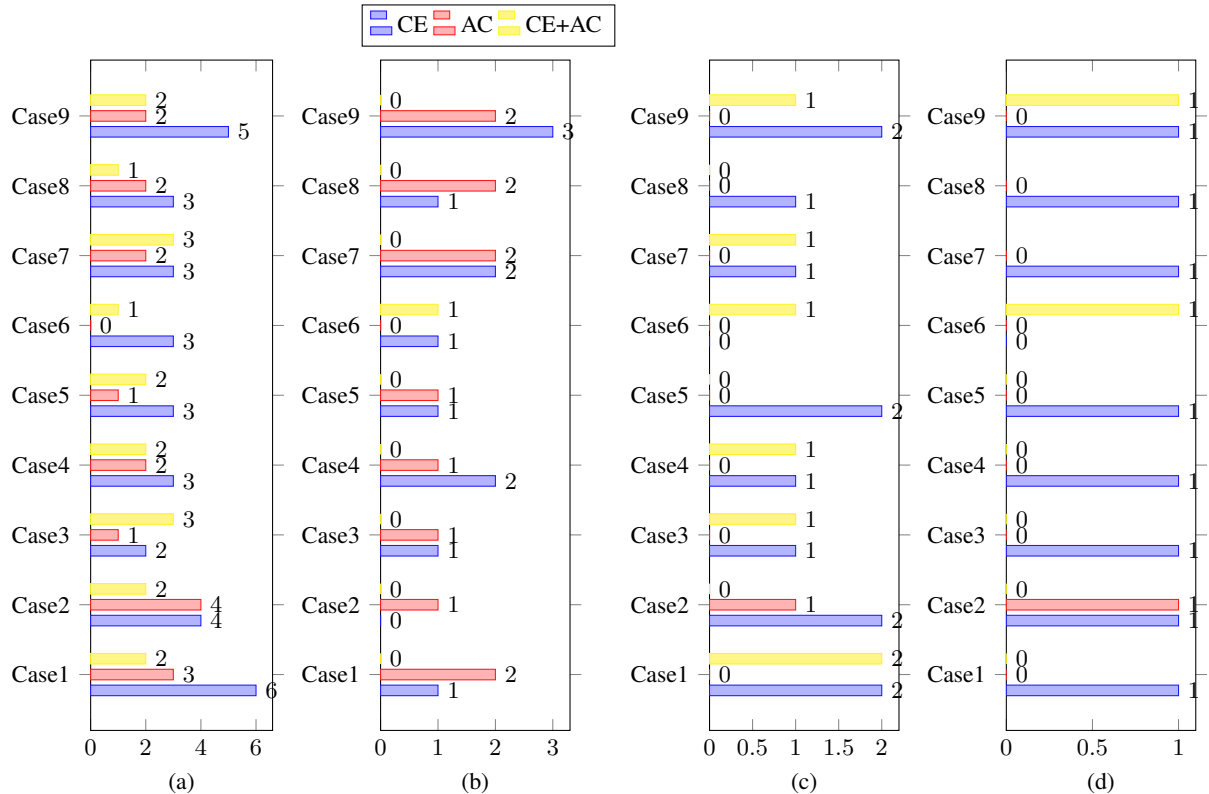
\begin{figure*}[ht]

 % \begin{centering}
%\begin{adjustwidth}{-1.5cm}{}
\begin{tikzpicture}  
\begin{axis}[legend style={at={(0.8,1.02)}, anchor=south west, legend columns=1},
    %tiny,
     xbar,
    xmin=0, 
    width=4cm, height=10.5cm, 
    xlabel={(a) },
    symbolic y coords={{Case1},{Case2}, {Case3},{Case4},{Case5},{Case6}, {Case7},{Case8},{Case9}},
    ytick=data,
    nodes near coords, nodes near coords align={horizontal},
    %every node near coord/.append style={font=\tiny},
]
 \addplot[fill=blue!30,draw=blue!90,tickwidth = 0pt,bar width=4pt,, tick label style={font=\tiny}] coordinates {
(6,{Case1}) 
(4,{Case2})
    (2,{Case3}) 
    (3,{Case4})
    (3,{Case5}) 
    (3,{Case6})
    (3,{Case7}) 
    (3,{Case8})
    (5,{Case9})
    };
    \addplot[fill=red!30,draw=red!90,tickwidth = 0pt,bar width=4pt,, tick label style={font=\tiny}] coordinates {
(3,{Case1}) 
    (4,{Case2})
    (1,{Case3}) 
    (2,{Case4})
    (1,{Case5}) 
    (0,{Case6})
    (2,{Case7}) 
    (2,{Case8})
    (2,{Case9})
    };
    \addplot[fill=yellow!60,draw=yellow!90,tickwidth = 0pt,bar width=4pt,, tick label style={font=\tiny}] coordinates {
(2,{Case1}) 
    (2,{Case2})
    (3,{Case3}) 
    (2,{Case4})
    (2,{Case5}) 
    (1,{Case6})
    (3,{Case7}) 
    (1,{Case8})
    (2,{Case9})
    };
    % \legend{\ac{CE}}
\end{axis}
\end{tikzpicture}  
\begin{tikzpicture}
\begin{axis}[legend style={at={(-0.1,1.02)}, anchor=south west, legend columns=3},
    %tiny,
     xbar,
    xmin=0, 
    width=4.1cm, height=10.5cm, 
    xlabel={(b)},
    symbolic y coords={{Case1},{Case2}, {Case3},{Case4},{Case5},{Case6}, {Case7},{Case8},{Case9}},
    ytick=data,
    nodes near coords, nodes near coords align={horizontal},
    %every node near coord/.append style={font=\tiny},
]
   \addplot[fill=blue!30,draw=blue!90,tickwidth = 0pt,bar width=4pt,, tick label style={font=\small}] coordinates {
(1,{Case1}) 
(0,{Case2})
    (1,{Case3}) 
    (2,{Case4})
    (1,{Case5}) 
    (1,{Case6})
    (2,{Case7}) 
    (1,{Case8})
    (3,{Case9})
    };
    \addplot[fill=red!30,draw=red!90,tickwidth = 0pt,bar width=4pt,, tick label style={font=\small}] coordinates {
(2,{Case1}) 
    (1,{Case2})
    (1,{Case3}) 
    (1,{Case4})
    (1,{Case5}) 
    (0,{Case6})
    (2,{Case7}) 
    (2,{Case8})
    (2,{Case9})
    };
    \addplot[fill=yellow!60,draw=yellow!90,tickwidth = 0pt,bar width=4pt,, tick label style={font=\small}] coordinates {
(0,{Case1}) 
    (0,{Case2})
    (0,{Case3}) 
    (0,{Case4})
    (0,{Case5}) 
    (1,{Case6})
    (0,{Case7}) 
    (0,{Case8})
    (0,{Case9})
    };
    \legend{\ac{CE},\ac{AC}, {CE}+{AC}}
\end{axis}
\end{tikzpicture}  
\begin{tikzpicture}
\begin{axis}[ legend style={at={(-1,1.02)}, anchor=south west, legend columns=1},
    %tiny,
     xbar,
    xmin=0,
    width=4.1cm, height=10.5cm,  
    xlabel={(c)},
    symbolic y coords={{Case1},{Case2}, {Case3},{Case4},{Case5},{Case6}, {Case7},{Case8},{Case9}},
    ytick=data,
    nodes near coords, nodes near coords align={horizontal},
    %every node near coord/.append style={font=\tiny},
]
   \addplot[fill=blue!30,draw=blue!90,tickwidth = 0pt,bar width=4pt,, tick label style={font=\small}] coordinates {
(2,{Case1}) 
(2,{Case2})
    (1,{Case3}) 
    (1,{Case4})
    (2,{Case5}) 
    (0,{Case6})
    (1,{Case7}) 
    (1,{Case8})
    (2,{Case9})
    };
    \addplot[fill=red!30,draw=red!90,tickwidth = 0pt,bar width=4pt,, tick label style={font=\small}] coordinates {
(0,{Case1}) 
    (1,{Case2})
    (0,{Case3}) 
    (0,{Case4})
    (0,{Case5}) 
    (0,{Case6})
    (0,{Case7}) 
    (0,{Case8})
    (0,{Case9})
    };
    \addplot[fill=yellow!60,draw=yellow!90,tickwidth = 0pt,bar width=4pt,, tick label style={font=\small}] coordinates {
(2,{Case1}) 
    (0,{Case2})
    (1,{Case3}) 
    (1,{Case4})
    (0,{Case5}) 
    (1,{Case6})
    (1,{Case7}) 
    (0,{Case8})
    (1,{Case9})
    };
     % \legend{,,Combination}
\end{axis}
\end{tikzpicture} 
\begin{tikzpicture}
\begin{axis}[ legend style={at={(-1,1.02)}, anchor=south west, legend columns=1},
    %tiny,
     xbar,
    xmin=0,
    width=4.1cm, height=10.5cm,  
    xlabel={(d)},
    symbolic y coords={{Case1},{Case2}, {Case3},{Case4},{Case5},{Case6}, {Case7},{Case8},{Case9}},
    ytick=data,
    nodes near coords, nodes near coords align={horizontal},
    %every node near coord/.append style={font=\tiny},
]
   \addplot[fill=blue!30,draw=blue!90,tickwidth = 0pt,bar width=4pt,, tick label style={font=\small}] coordinates {
(1,{Case1}) 
(1,{Case2})
    (1,{Case3}) 
    (1,{Case4})
    (1,{Case5}) 
    (0,{Case6})
    (1,{Case7}) 
    (1,{Case8})
    (1,{Case9})
    };
    \addplot[fill=red!30,draw=red!90,tickwidth = 0pt,bar width=4pt,, tick label style={font=\small}] coordinates {
(0,{Case1}) 
    (1,{Case2})
    (0,{Case3}) 
    (0,{Case4})
    (0,{Case5}) 
    (0,{Case6})
    (0,{Case7}) 
    (0,{Case8})
    (0,{Case9})
    };
    \addplot[fill=yellow!60,draw=yellow!90,tickwidth = 0pt,bar width=4pt,, tick label style={font=\small}] coordinates {
(0,{Case1}) 
    (0,{Case2})
    (0,{Case3}) 
    (0,{Case4})
    (0,{Case5}) 
    (1,{Case6})
    (,{Case7}) 
    (,{Case8})
    (1,{Case9})
    };
     % \legend{,,Combination}
\end{axis}
\end{tikzpicture} 
\caption{\centering  Security Controls Inhibition Pattern for Ransomware Cases -- (a) Security Controls appeared in all level of \ac{IFT}, (b) Security Controls at bottom level (L1) of \ac{IFT}, (c) Security Controls at Phase1 (P1) of \ac{IFT}, (d) Security Controls at L1P1 of \ac{IFT}
}
%\end{adjustwidth}
\label{fig:ransomware_findings}
 % \end{centering}
\end{figure*}

We investigated 9 cases of ransomware attacks, where the variants included Black Basta, Conti, Lockbit among others. Our analysis was conducted in two phases. The first phase involved determining the overall controls employed, including both \ac{CE} and \ac{AC}, as well as the combination of both, to prevent the attacks going further in a tree. In the second phase, we performed detailed analysis and examined the controls at L1, P1 and L1P1 to determine which controls were effective in halting the progression of events in the \ac{IFT}. Our findings are presented in Figure \ref{fig:ransomware_findings}.  The blue, red and yellow bar represents the \ac{CE} controls, \ac{AC} and combination of both controls respectively.

Figure \ref{fig:ransomware_findings} (a) displays the distribution of controls utilised in the different ransomware incidents.  In cases 1 and 9 it is observed that \ac{CE} controls alone mitigate the majority of attack pathways. The phenomenon is the same for cases 4, 5, 6, 7 \& 8 where \ac{CE} controls by themselves are appropriate to mitigate most attack pathways, however \ac{AC} and combination of \ac{AC} \& \ac{CE} controls are applicable to mitigate the remaining attack pathways. On the other hand, for case 2, \ac{CE} \& \ac{AC} controls are independently applicable to mitigate an equal number of attack pathways. Case 3 is an exception, where our findings show that a combination of \ac{CE} \& \ac{AC} are suitable to mitigate more attack pathways than they do when applied independently.

 Figure \ref{fig:ransomware_findings} (b) presents the effectiveness of controls from preventing an attack if proceed beyond L1; we find that \ac{CE} controls can be applied to stop a ransomware attack in cases 4 \& 9. In our investigations we find the applicability of a combination of \ac{CE} \& \ac{AC} in case 6 only to prevent an attack at L1. However, for all the other instances of ransomware attack we find that application of \ac{AC} to be effective at L1. 

 The phase analysis as presented in Figure \ref{fig:ransomware_findings} (c) shows that \ac{CE} by themselves are appropriate to prevent more attack pathways at the initial phase but they need to be supported with a combination of \ac{CE} \& \ac{AC} controls. For cases 1, 3, 4, \& 7, \ac{CE} controls and a combination of \ac{CE} \& \ac{AC} are evenly poised in their ability to prevent attack pathways at the initial phase. 

Somewhat of a contrast emerges when we combine L1 and P1. Figure \ref{fig:ransomware_findings} (d) shows that \ac{CE} controls by themselves can be applied to prevent all the attack pathways in 6 of the 9 incidents of ransomware attack. Case 6 is a departure, where \ac{AC} are applicable to prevent all the attack pathways and for cases 2 \& 9, we need to apply \ac{AC} and a combination of \ac{CE} \& \ac{AC} respectively. 

While looking at specific controls (Table \ref{table:ransomware_pattern}), we find that \ac{CE} controls such as Firewall and Secure Configuration were overwhelmingly applicable. For all the ransomware incidents Secure Configuration was the most commonly applicable \ac{CE} control, followed by Malware Protection. So far as the \ac{AC}, we find Encryption and Backup were widely applicable.

% To inhibit the attack pathways, secure configuration is the commonly used lower level \ac{CE} control in Black Basta and Conti ransomware. On the other hand, security updates and malware protection are both L1 controls in LockBit and user access control in other types of ransomware. The most commonly used lower level \ac{AC} are backup, encryption and policy.

\begin{table}[ht]
\scriptsize
\centering
 \begin{tabular}{p{3cm} |p{3.5cm} |p{3cm} |p{3cm}} 
 \hline
 \textbf{Ransomware Variant} & \textbf{Most Used CE} & \textbf{Most Used AC} & \textbf{Combination Controls}  \\ [0.5ex] 
 \hline

 Black Basta & -Firewall & -Education & -Malware Protection + Education \\
            &   -Secure Configuration (L1)    & -Encryption &  \\
            & & -Backup (L1) &\\\hline
Conti       & -Secure Configuration (L1) & -Education & -Secure Configuration + Education \\
            &  -Security Update Management      & -Backup      &  -Backup + Firewall\\
            &-Malware Protection& -Encryption (L1)&\\\hline
 LockBit & -Security Update Management (L1) & -Backup (L1) & -Secure Configuration + Policy \\
            &  -Malware Protection (L1)      & -Encryption & -Logging \& Monitoring + User Access Control \\
            && -Policy&\\\hline
 Others & -Secure Configuration& -Backup (L1) & -Secure Configuration + Logging \& Monitoring \\
         & -User Access Control (L1)& -Logging \& Monitoring & -Secure Update Management + Education\\
         && -Policy (L1)&\\[1ex] 
 \hline
 \end{tabular}
 \caption{Inhibit Pattern in Ransomware Attacks. L1 represent the layer1 controls}
 \label{table:ransomware_pattern}
\end{table}

\section{Discussion}\label{discussion}

\subsection{Need for Additional Controls}
\ac{CE} were designed as a set of usable \& affordable controls intended to be used by organisations. They focus on patch management, access control, malware protection, firewall and secure configuration. The proposed controls point to basic hygiene that an organisation should practice to prevent security breaches. The assumption is that attackers are opportunistic, and all attackers do not come with formidable expertise. 

We studied the attack scenarios by phases, levels, and a combination of the two, along with an overall analysis of the entire attack lifecycle. While CE controls seem to be effective in preventing many malicious attack pathways from propagating beyond the initial phases \& levels, there remains a need for \ac{AC} to mitigate the attacks when they progress further into the systems and networks. Furthermore, the need for \ac{AC} cannot be completely ruled out even at the initial stages. We identify some \ac{AC} - encryption, backup, policy, education, logging \& monitoring. Consequently, we see two distinct lines of deliberations here. One is on the need for \ac{AC}; the other being, should there be a need for \ac{AC}, then deliberating on the underpinning knowledge for a \ac{CE} assessor. 

Considering \ac{AC} includes a fundamental debate as to whether controls should only be preventive and not include mechanisms that help in recovery. On the other hand, our results show that while attacks propagate beyond the initial levels and phases there is indeed a need for \ac{AC}. The argument for \ac{AC} can be grounded upon the idea that it is difficult to anticipate all possible attack pathways and harden systems such that attacks can be prevented from progressing beyond the initial stages. With respect to the need for recovery mechanisms, there are applications where recovery is integral to intrusion and fault tolerance~\cite{castro2002practical} and an important ingredient of cyber resilience~\cite{ullah2018data}. Protection mechanisms need to evolve across space and time~\cite{chowdhury2023threat}. 

Should there be \ac{AC} as part of CE, the other pertinent question is redesigning the syllabus for a \ac{CE} assessor. Prior research conducted a systematic mapping with CyBOK and found that \emph{cloud security} and \emph{applied cryptography} covers the skills that an assessor needs. Based on our analysis, a further expansion of the syllabus could include the \emph{human factors} and \emph{security operations and incident management} Knowledge Areas (these are already partly covered in the syllabus, albeit not extensively). Future research could consider mapping the ingredients of the \ac{AC} to elicit relevant Knowledge Areas.

\begin{finding}
    CE controls are effective in preventing attacks at their inception, but need the support of Additional Controls if attacks penetrate further into an organisation's systems \& networks. Deliberations on \ac{AC} spawn a foundational debate as to whether controls should only be pre-emptive and exclude recovery from breaches. Should there be recognition of \ac{AC} as essential then \ac{CE} assessors need to be skilled in \emph{human factors} and \emph{security operations and incident management}.
\end{finding}

\subsection{Deployment of the controls}
The \emph{inhibit gates} indicate controls that can protect against breaches expanding within an organisation. Our investigations reveal that the deployment of controls needs to be contextualised. To that we propose two possibilities for effective deployment of controls. 

\begin{figure}[ht]
        \centering
        \includegraphics[scale=0.70]{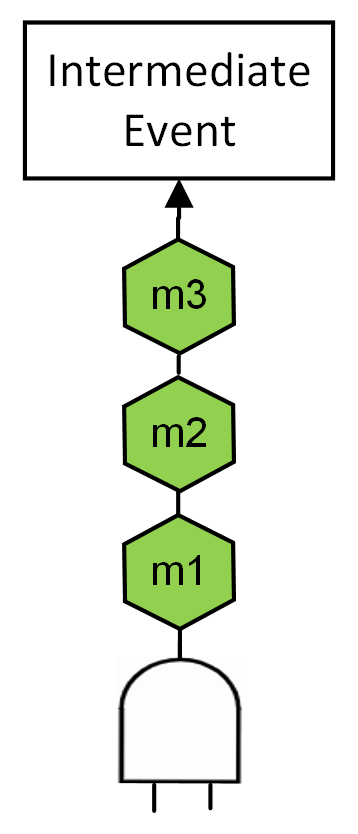}
        \caption{Sequential Controls}
        \label{fig:seq}
        \end{figure}

\paragraph{Sequential Controls}  
       Inhibit gates refer to the controls that must be activated in a specific order to mitigate the particular event effectively. These controls should be implemented in a specific sequence, where each inhibit gate depends on the previous one to work successfully. For example, in a phishing attack which is trying to perform privilege escalation, the  inhibit gates involve User Education (m1), Multi-factor Authentication (m2) and Monitoring the logging attempts (m3). These three controls should be in the sequence (m1-m2-m3) as shown in Figure \ref{fig:seq} and executed correctly; otherwise, overall cyber controls could fail.

\begin{figure}[ht]
        \centering
        \includegraphics[scale=0.70]{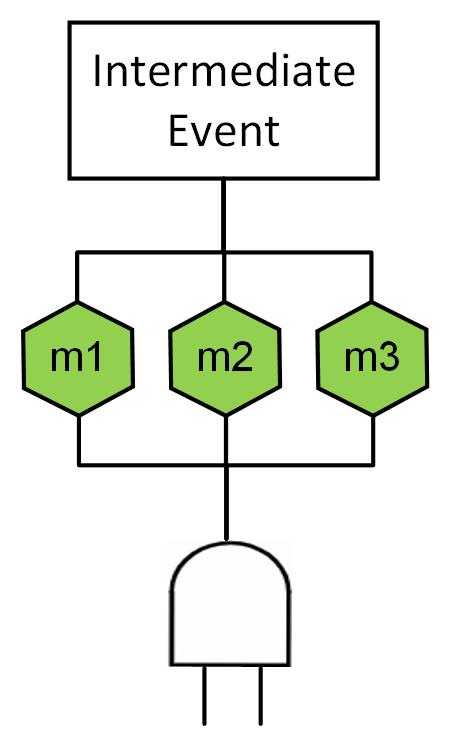}
        \caption{Parallel Controls}
        \label{fig:parallel}
        \end{figure}

\paragraph{Parallel controls} Parallel inhibit gates work independently and ensure that if one control fails, other controls are in place and sufficient to prevent the attack (Figure \ref{fig:parallel}). For example, in the case of
unauthorised access to an account, the most common used security controls, such as user access control (m1), multi-factor Authentication (m2) and firewall (m3), can be used in parallel (m1 $\parallel$ m2 $\parallel$ m3)  for attack mitigation. If one of the controls fails, the other can still provide some level of protection.

\begin{finding}
    Our analyses show that \ac{CE} controls, when deployed in conjunction with other controls for certain contexts, can provide improved deterrence and defence. This means controls can be sequentially placed or placed in parallel to another. Should one of them fail the other can still mitigate against certain attacks. For example, we can situate difficult to measure Additional Controls such as \emph{awareness \& education} along with multi-factor authentication to protect against certain types of phishing attacks. 
\end{finding}

\subsection{Comparative understanding with previous work}
There are prior investigations into the effectiveness of Cyber Essential controls. The corpus of attacks we investigated is smaller than~\cite{lancasterreport}. However, the cases we investigate are detailed incidents that provide an in-depth view of the controls needed to prevent attacks. Our analysis shows that CE controls are effective in preventing most of the attacks in our dataset at the initial stages. However, a comparative understanding of the incident reports between~\cite{lancasterreport} and ours reveals that while Cyber Essential controls are robust against malicious remote access, like establishing command \& control channels and remote desktops, they do not mitigate against human factors, especially during attack staging. The impact of the latter is pronounced, as we see in our corpus that includes ransomware incidents. If attacks progress beyond the initial stages there is a more pronounced need for \ac{AC}, and \ac{CE} controls by themselves are not enough to protect against them. There is a synergy between the additional socio-technical \& technical controls suggested by prior work~\cite{lancasterreport} and our work. Table \ref{table:table4} presents the suggested \ac{AC} across prior reports and our findings. Logging \& monitoring, as a technical control is suggested by the prior studies~\cite{lancasterreport}, a control suggested by our investigations as well. Awareness \& education is a socio-technical control suggested by the previous reports as well as in our work.  

A contrasting case is of patch management between the previous study and ours. We identify patch management as an effective control for 19 of the 45 incidents which is roughly \(42\%\) compared to \(87.5\%\) in the previous study. % Future investigations can consider a much larger corpus to further investigate this shift. 
Firewalls, on the other hand, are identified in \(90\%\) of the incidents we investigated in our report.  We do not see emphasis on back-ups in the previous report which we identify as an additional control - this might be due to both the increase in ransomware since 2015 and our specific focus on  ransomware incidents as an element of our investigation. 

The previous study reports that \ac{CE} technical controls cannot mitigate threats which are due to design decisions (they articulate as hard coded flaws) in the hardware and software. A scrutiny of our findings reveals that attackers take advantage of the enablers which we indicate through \emph{basic gates}. These enablers reflect the manner in which systems are configured, for example, PowerShell permissions. While the fundamentals of protection mechanisms are there~\cite{saltzer1975protection}, developers come with diverse security understandings. Conventional recommendations point towards designing software with security in mind~\cite{mcgraw1999software}. Future research could explore the extent to which security can be woven into deployment practices, with explicit focus on better observance of the principles.

\begin{finding}
    A comparative understanding with prior work shows that \ac{CE} controls remain resilient against many attacks. However, our analysis and prior work points to the need for additional controls for improved mitigation. Controls around \emph{human factors} and \emph{logging \& monitoring} are found by both the studies. While prior work argues for safe design defaults in software \& hardware, we make a case for safe software deployment defaults.  
\end{finding}

\begin{table}[ht]
\centering
\begin{tabular}{ |p{1.8cm}|p{1.7cm}|p{2cm}|p{2cm}|
}
 \hline
 \multicolumn{4}{|c|}{\textbf{\normalsize Sociotechnical Controls}} \\
 \hline

 \centering
 \textbf{Additional Controls} & \textbf{Findings in our Report} & \textbf{Findings in CyBoK Report} & \textbf{Findings in Lancaster Report}\\
 \hline
\centering
Awareness \& Education   &  \centering \checkmark   &  \centering \checkmark &  \checkmark  \\ \hline
\centering Policy & \centering  \checkmark  & \centering  \checkmark   & -\\
\hline

\multicolumn{4}{|c|}{\textbf{\normalsize	 Technical Controls}} \\
\hline
\centering  Logging \& Monitoring & \centering \checkmark &\centering \checkmark & \checkmark \\ \hline
 \centering Backup    & \centering \checkmark & \centering \checkmark &  -\\ \hline
 \centering Encryption  &  \centering \checkmark   & \centering \checkmark & - \\
 \hline
\end{tabular}
\caption{Mapping Findings with Previous Research}
    \label{table:table4}
\end{table}

\section{Conclusion}\label{conclusion}
Security is a property of technology with a defined objective. \ac{CE} controls come with defined objectives of preventing commodity attacks and are meant for a wide range of users. We explore the effectiveness of \ac{CE} controls using 45 incident reports, which includes 9 incidents of ransomware. \ac{IFT} allow us the granularity to break incidents into non-reducible events and build upon them to construct the attack pathways across levels and phases.  The nuance offered by the method allows us to explore if attacks can be prevented right at their incursion. For example, our level and phase analysis shows that \ac{CE} alone can completely mitigate 25 out of the 45 incidents at their initial phases, while using a combination of \ac{CE} and Additional Controls can mitigate another 19 incidents. The additional controls we propose as part of this work are \emph{education, logging \& monitoring, policy, encryption} \& \emph{back up}. The insights from our overall study can be summarised as:

\begin{itemize}

\item \ac{CE} controls by themselves are effective against the majority of commodity attacks, however, they need to be complemented with additional controls for complete mitigation of all the attacks we studied. A combination of both, reduces the attack surface and consequently the blast radius. For example, the combined effectiveness of \ac{CE} and \ac{AC} prevent 44 of 45 attacks at their initial phases thus confining the blast radius, if any. 

\item Security ought to expand beyond the paradigm of prevention to recovery. This is more attuned to the ever-evolving dynamic threat landscape. The need to do so has been recognised earlier as well. For example, \emph{back-up} is recommended as a complementary good practice in \ac{CE} guidance to small businesses. 

\item We find synergy between our recommendations with those from prior work in this area. Logging \& monitoring and human factors  are among the common recommendations. Our results show the importance safe software deployment defaults along with safe design defaults.  

\item Security controls around human factors are difficult to measure and thus difficult to include as \ac{CE} controls. We would like to build upon insights from human centered security research over the last few decades. Systems that fail to evolve with human needs are not only difficult to use but fail to meet the legitimate security expectations of their users.

\end{itemize} 

\bibliography{main}
\bibliographystyle{plain}

% \begin{appendices}
% \section{Over Controls Utilised to Inhibit Attack Pathways of an IFT for Each Cases}
% % \clearpage
% \begin{figure*}
%      \centering
%     \includegraphics[scale=0.3]{t1.png}
%     \includegraphics[scale=0.3]{t2.png}
%     \includegraphics[scale=0.3]{t3.png}
%     \caption{Overall CE and AC Controls utilised in all 45 corpus of cyber incidents}
%     \label{fig:my}
% \end{figure*}
% \end{appendices}

% \appendices
% \newpage

\appendix
    \clearpage
    \section{Over Controls Utilised to Inhibit Attack Pathways of an IFT for Each Cases}    
    \begin{figure}[ht]
         \centering
        \includegraphics[scale=0.35]{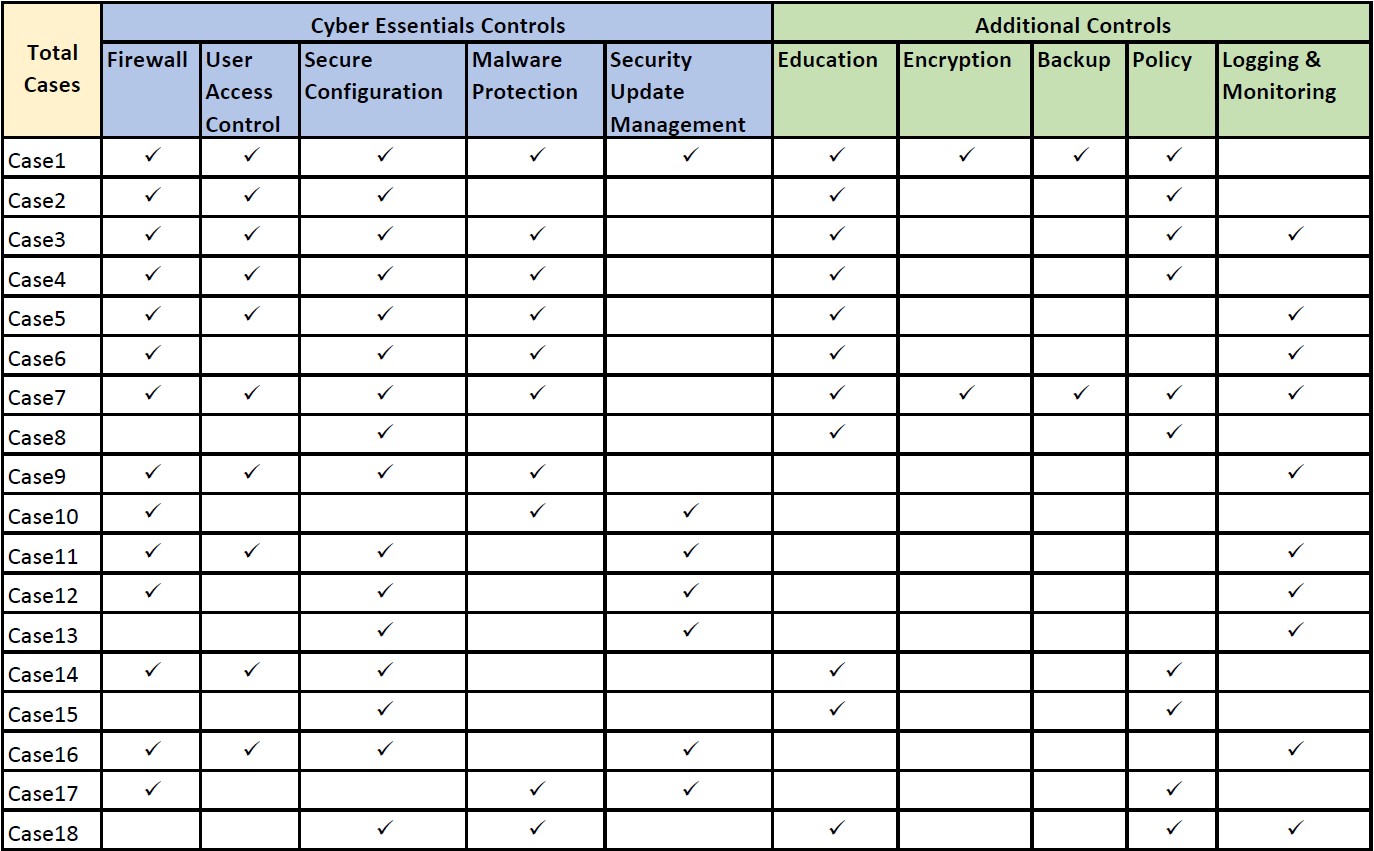}
        \includegraphics[scale=0.35]{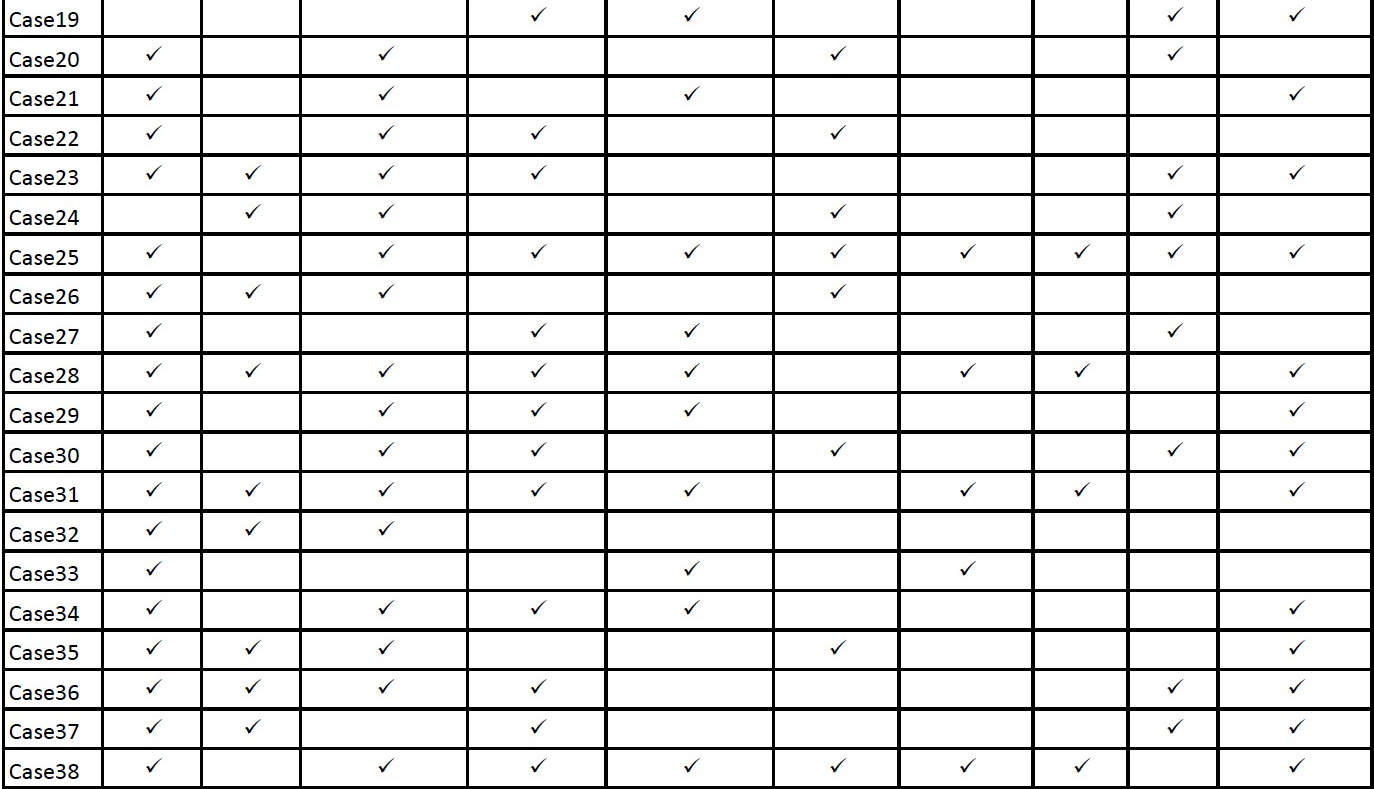}
        \includegraphics[scale=0.35]{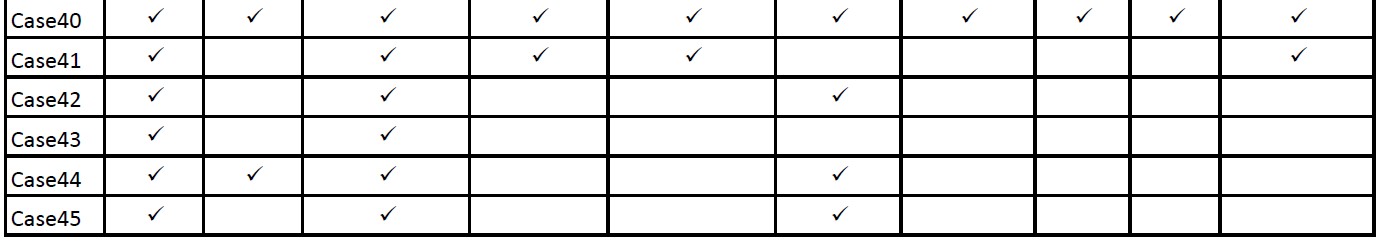}
        \caption{Overall CE and AC Controls utilised in all 45 corpus of cyber incidents}
        \label{fig:my}
        \end{figure}

\end{document}